\documentclass[11pt]{article}

\usepackage{amsmath}
\usepackage{amsfonts}
\usepackage{amssymb}
\usepackage{latexsym}
\usepackage{enumerate}
\usepackage{graphicx}
\usepackage[english]{babel}

\newcommand{\captionfonts}{\small}

\makeatletter  
\long\def\@makecaption#1#2{%
  \vskip\abovecaptionskip
  \sbox\@tempboxa{{\captionfonts #1: #2}}%
 \ifdim \wd\@tempboxa >\hsize
    {\captionfonts #1: #2\par}
  \else
    \hbox to\hsize{\hfil\box\@tempboxa\hfil}%
  \fi
  \vskip\belowcaptionskip}
\makeatother   

\usepackage[
      colorlinks=true,
      linkcolor=black,
      urlcolor=blue,
      filecolor=blue,
      citecolor=blue,
      pdfstartview=FitH,
      pdftitle={},
      pdfauthor={},
      pdfsubject={},
      pdfkeywords={},
      pdfpagemode=UseNone,
      bookmarksopen=true
      ]{hyperref}
\usepackage[all]{hypcap}     

\begin{document}
\input epsf

\def\p{\partial}
\def\h{{1\over 2}}
\def\be{\begin{equation}}
\def\bea{\begin{eqnarray}}
\def\ee{\end{equation}}
\def\eea{\end{eqnarray}}
\def\d{\partial}
\def\la{\lambda}
\def\eps{\epsilon}
\def\bb{\bigskip}
\def\mm{\medskip}
\newcommand{\dm}{\begin{displaymath}}
\newcommand{\edm}{\end{displaymath}}
\renewcommand{\b}{\tilde{B}}
\newcommand{\gm}{\Gamma}
\newcommand{\ac}[2]{\ensuremath{\{ #1, #2 \}}}
\renewcommand{\ell}{l}
\newcommand{\z}{\ell}
\newcommand{\newsection}[1]{\section{#1} \setcounter{equation}{0}}
\def\bb{$\bullet$}
\def\Qbar{{\bar Q}_1}
\def\QPbar{{\bar Q}_p}

\def\q{\quad}

\def\bn{B_\circ}

\let\a=\alpha \let\b=\beta \let\g=\gamma \let\d=\delta \let\e=\epsilon
\let\c=\chi \let\th=\theta  \let\k=\kappa
\let\l=\lambda \let\m=\mu \let\n=\nu \let\x=\xi \let\r=\rho
\let\s=\sigma \let\t=\tau
\let\vp=\varphi \let\vep=\varepsilon
\let\w=\omega      \let\G=\Gamma \let\D=\Delta \let\Th=\Theta
                     \let\P=\Pi \let\S=\Sigma

\def\h{{1\over 2}}
\def\t{\tilde}
\def\r{\rightarrow}
\def\nn{\nonumber\\}
\let\bm=\bibitem
\def\Kt{{\tilde K}}
\def\b{\vspace{3mm}}
\def\sq{{1\over \sqrt{2}}}

\let\p=\partial

\begin{flushright}
\end{flushright}
\vspace{20mm}
\begin{center}
{\LARGE  The flaw in the firewall argument}
\\
\vspace{18mm}
{\bf Samir D. Mathur ~and~ David Turton} \\

\vspace{15mm}
Department of Physics,\\ The Ohio State University,\\ Columbus,
OH 43210, USA

\vspace {5mm}
mathur.16@osu.edu,~turton.7@osu.edu
\vspace{4mm}

\end{center}
\vspace{10mm}

\thispagestyle{empty}
\begin{abstract}

\vspace{6mm}

 A lot of confusion surrounds the issue of black hole complementarity, because the question has been considered without discussing the mechanism which guarantees unitarity.  Considering such a mechanism leads to the following: (1) The Hawking quanta with energy $E$ of order the black hole temperature $T$ carry information, and so only appropriate processes involving $E\gg T$ quanta can have any possible complementary description with an information-free horizon; (2) The stretched horizon describes all possible black hole states with  a given mass $M$, and it must expand out to a distance $s_{bubble}$ before it can accept additional infalling bits; (3) The Hawking radiation has a specific low temperature $T$, and infalling quanta interact significantly with it only within a distance $s_\alpha$ of the horizon. One finds $s_\alpha\ll s_{bubble}$ for $E\gg T$, and this removes the argument against complementarity recently made by Almheiri et al. In particular, the condition $E\gg T$ leads to the notion of `fuzzball complementarity', where the modes around the horizon are indeed correctly entangled in the complementary picture to give the vacuum.

\end{abstract}

\vspace{8mm}

\vskip 1.0 true in

\newpage

\numberwithin{equation}{section} 

\setcounter{tocdepth}{1}
\tableofcontents

\baselineskip=14pt
\parskip=3pt

\section{Introduction}\label{intro}

Hawking's discovery of black hole evaporation led to a sharp puzzle \cite{hawking}. Pairs of particles are created at the horizon, with one member of the pair $b$ escaping to infinity as radiation, and the other member $c$ falling into the hole  and reducing its mass. The crucial point is that $b$ and $c$ are entangled, so the entanglement entropy $S_{ent}$ of the emitted radiation $\{ b_i\}$ with the remaining hole keeps rising till 
we reach near the endpoint of evaporation. This is in sharp contrast to the behavior of a normal body, where $S_{ent}$ starts to reduce after the halfway point and reaches zero at the end of the evaporation process \cite{page}. 

Hawking's argument was very robust because it used only the assumption that
the region around the horizon was a piece of `normal' spacetime. The result was recently shown to hold even when small corrections to the leading order process are included \cite{cern}.
More explicitly, the result assumes that (i) the state at the horizon is close to the vacuum $|0\rangle$ that we have in the lab, and (ii) the evolution Hamiltonian is also close to the laboratory Hamiltonian $H_{lab}$. Thus we may state Hawking's result as
\begin{quote}
H:~~~If there is no nontrivial structure at the horizon, then  $S_{ent}$  will keep rising until 
near the endpoint of evaporation; it will not start reducing at the half-way point.
\end{quote}
We can turn this around and write an exactly equivalent statement of Hawking's result:
\begin{quote}
H':~~~If we want  $S_{ent}$  to  start reducing at some point, then we must have nontrivial structure at the horizon.
\end{quote}

There has been a lot of interest, as well as quite some confusion,  generated by a recent paper by Almheiri, Marolf, Polchinski and Sully  (AMPS)~\cite{amps} on whether it is possible to smoothly fall into black holes.\footnote{ For related earlier work see~\cite{Braunstein:2009my}, and for a more complete list of references see~\cite{ampss}.}
In the literature that followed, it appears that many people confused the AMPS result with Hawking's original statement H'. The difficulty that people had in arguing around the AMPS argument was then the same as the difficulty that people had all these years in resolving Hawking's paradox. This confusion, of course, does AMPS no service, since it would subsume their argument into Hawking's original statement. So we will start by removing this confusion, and then proceed in the following steps:

\b

(i) We recall the initial formulation of complementarity \cite{complementarity}, which we denote `traditional complementarity'. This notion was formulated {\it before} we understood how the information paradox was to be resolved, and involved postulating new physics in the form of a special  kind of nonlocality.  We recall the AMPS argument against this `traditional complementarity'. 

\b

(ii) In recent years we have understood that the information paradox is solved through the fuzzball construction in string theory, without invoking new physics beyond standard string theory (see \cite{reviews,beyond} for reviews). This gives real rather than virtual degrees of freedom at the horizon. A notion of complementarity was developed for this situation in \cite{plumberg,beyond}; this conjecture was denoted `fuzzball complementarity' in~\cite{mt}. This complementarity is based on a crucial approximation $E\gg T$ which AMPS do not consider; thus their argument fails to address fuzzball complementarity from the start. Here $E$ is the energy relevant to the physical process, and $T$ is the Hawking temperature, both measured at infinity. 

\b

(iii) One might try to extend the AMPS argument to fuzzball complementarity, but here one finds that AMPS miss the issues important in such an analysis:

\vspace{1mm}

(a) When a quantum with $E\gg T$ falls from afar onto the stretched horizon, a large number of {\it new} degrees of freedom are created; these new degrees of freedom are not entangled with anything, and it is {\it their} dynamics which is conjectured to be captured by the complementary description.

\vspace{1mm}

(b) We expect that these new degrees of freedom are accessible when the infalling quantum reaches the stretched horizon.  AMPS assume that the stretched horizon does not respond before the infalling quantum actually reaches it. For example, consider a Schwarzschild black hole of mass $M$. The stretched horizon is a very special surface: its possible states exhaust all the $Exp[S_{bek}]$ states that are possible within area $A=A(M)$, where $S_{bek}$ is the Bekenstein-Hawking entropy. If we could place a new quantum on this surface without first expanding the surface, then we would have {\it more} than $Exp[S_{bek}]$ states on a surface with area $A(M)$; in contradiction with what we expect from the stretched horizon. Thus the stretched horizon should expand out {\it before} it is hit by the infalling quantum. With fuzzballs, the tunneling mechanism that solves the information problem \cite{tunnel, rate} leads to the required expansion of the stretched horizon, so the new degrees of freedom are indeed accessed {\it before} the time that AMPS would consider in their analysis.

\vspace{1mm}

(c) One might be concerned that interaction with outgoing Hawking quanta might scatter an infalling  quantum into a new state before the new degrees of freedom are accessed; in this case the excitation of the new degrees of freedom would not capture the state of the infalling quantum.  But Hawking radiation is radiation with a  very particular (low) temperature $T$. We estimate the required scattering cross section for $E\gg T$ and show that  it is small at the location where the new degrees of freedom are accessed.  Thus the infalling quantum does {\it not} get destroyed by Hawking radiation before the new (unentangled) degrees of freedom are accessible. As a result, the AMPS argument fails to rule out the existence of a complementary description.


In short, the discussion around AMPS has missed the idea of how complementarity is to be obtained. AMPS observe that Hawking modes with energy $E\sim T$ provide a nontrivial structure near the horizon, since they are not in the vacuum state. As we have noted above, this is guaranteed to be the case by Hawking's statement H'. AMPS then worry that an infalling observer cannot avoid interacting with these modes, so he will get `burnt'. But here they are asking the wrong question. The point is {\it not} to look for a way to avoid interaction with this structure near the horizon.  Rather, what we are looking for is that an infalling quantum `smash' onto this structure, and create excitations, in the same way that a graviton falling onto a stack of D-branes `smashes' and creates excitations of gluons. The gluons evolve in a way that can be duplicated by free infall into AdS space, and we can ask if smashing onto the structure at the black hole horizon is a process with a similar `complementary' description. When we ask the question this way, we come across the above issues (a)-(c), and find that the AMPS argument does not rule  out the conjecture of fuzzball complementarity.

\section{Summary of the issues and the proposal of fuzzball complementarity}

The discussion of complementarity becomes confusing if it is not accompanied by a discussion of how the information problem is solved in the theory under consideration. In this section we delineate the relevant issues: the information paradox, the idea of `traditional complementarity' and the idea of `fuzzball complementarity'.

\subsection{Structure at the horizon}\label{structure}

Let us now return to our discussion of Hawking's original result; as we had noted above, some of the literature following AMPS has confused Hawking's result with the AMPS argument.  We can further split Hawking's statement H' into two possibilities:

\b

(a) We keep the state $|0\rangle$ but alter the dynamics by assuming (hitherto unknown) nonlocal effects. Then the vacuum $|0\rangle$ will still produce entangled pairs, but nonlocal effects can remove the entanglement at a later stage (see for example \cite{giddings}). AMPS remark that this possibility leads to awkward effects, which is of course true; all familiar physics arises from local Lagrangians. 

\b

(b) We alter the state $|0\rangle$ but keep the dynamics local. Note that in this case the vacuum must be altered in modes down to the Planck length, not just at length scales of order the horizon radius $r_0$. This follows because if the wavemodes of order $l_p\ll \lambda \ll r_0$ were `normal' then they would evolve to make entangled pairs (since we have now assumed that evolution is normal at scales $l_p\ll\lambda$). Then we would again need nonlocality to remove the entanglement. Here `normal' refers to ordinary lab physics: evolution of long wavelength modes ($\lambda\gg l_p$) is given by local quantum field theory on curved spacetime, with corrections controlled by some small parameter $\epsilon$. These corrections may come from any quantum gravity effect, and all we require is that $\epsilon\r 0$ as $M\r \infty$, where $M$ is the mass of the black hole. 

In particular, since the main AMPS argument does not consider nonlocal effects, they are automatically led to a situation where the vacuum is corrupted at all length scales. 

\b

A priori, there is a third possibility that must be ruled out first: invalidation of the Hawking argument by `accumulation of small corrections'. We discuss this next, as the required argument \cite{cern} provides the setup for the AMPS discussion.

\subsection{Small corrections and strong subadditivity}\label{sec:cern}

In string theory we do not know of any long-distance nonlocal effects, so if we wish to have $S_{ent}$ decrease at any point then we need to have a state other than $|0\rangle$ at the horizon; i.e., we need black hole `hair'. This is of course the crux of the information paradox: if one can find hair, then the state at the horizon is not $|0\rangle$, and Hawking's argument will fail. String theorists did not have, until recently, a construction of this hair, but many of them were still not worried about Hawking's paradox. The reason was based on the following misconception. Suppose the horizon was a place with `normal physics', and let us include a {\it small} correction, order $\epsilon\ll 1$ to the state of each created pair. The number of pairs $N$ is very large, so it might be that suitable choices of these small corrections would lead to a situation where $S_{ent}$ does decrease in the manner expected of a normal body. 

A priori, it is not wrong to  think that small corrections might cause $S_{ent}$ to decrease.  Suppose the entangled pair at the first step is $\sq(|0\rangle_{b_1}|0\rangle_{c_1}+|1\rangle_{b_1}|1\rangle_{c_1})$. At the next step we can have the state
\bea
|\Psi\rangle&=&\h\Big ( |0\rangle_{b_1}|0\rangle_{c_1}[(1+\epsilon_1)|0\rangle_{b_2}|0\rangle_{c_2}+(1-\epsilon_1)|1\rangle_{b_2}|1\rangle_{c_2}]\nn
&&~~~~~~~~+|1\rangle_{b_1}|1\rangle_{c_1}[(1+\epsilon'_1)|0\rangle_{b_2}|0\rangle_{c_2}+(1-\epsilon'_1)|1\rangle_{b_2}|1\rangle_{c_2}] \Big )
\eea
Note that the correction at each step can depend on everything in the hole at all earlier steps; the only requirement is that the correction be small:  $|\epsilon_1|<\epsilon$, $|\epsilon'_1|<\epsilon$. 
We have $\sim 2^N$ correction terms in general after $N$ steps. Since $N\sim ({M\over m_p})^2 $ for a 3+1 dimensional black hole, it appears  a priori possible for small corrections to pile up to make $S_{ent}$ decrease after the halfway point of evaporation. 

In \cite{cern} it was proved, using strong subadditivity,  that such small corrections {\it cannot} lead to a decrease in $S_{ent}$. AMPS invoked this argument in their analysis, so let us outline the steps in \cite{cern}. Let $\{ b_1, \dots b_N\} \equiv \{ b_i\}$ be the quanta radiated in the first $N$ steps, and $\{ c_i\}$ their entangled partners. The entanglement entropy at step $N$ is $S_{ent}(N)=S(\{ b_i\})$. The created quanta at the next step are $b_{N+1}, c_{N+1}$. We then have \cite{cern}:

(i) By direct computation, one obtains
\be
S(b_{N+1}+c_{N+1})<\epsilon \,.
\label{pair}
\ee

(ii) Similarly, by direct computation one obtains
\be
S(c_{N+1})>\ln 2-\epsilon \,.
\label{pairq}
\ee

(iii) The unitary evolution of the hole does not affect quanta already emitted (we have assumed that nonlocal effects, if any extend only to distances of order $r_0$, and thus do not affect quanta that have been emitted from the hole long ago). Thus we have
\be
S(\{(b_i\})=S_N \,.
\ee

(iv) The strong subadditivity inequality gives
\be 
S(\{ b_i\}+b_{N+1})+S(b_{N+1}+c_{N+1})\ge S(\{ b_i\})+S(c_{N+1}) \,.
\ee
Using (i)-(iii) above we find that the entanglement entropy of the radiation after the $(N+1)$-th time step, $S_{N+1}\equiv S(\{ b_i\}+b_{N+1})$, satisfies
\be
S_{N+1}>S_N+\ln 2 -2\epsilon \,.
\label{inequality}
\ee
Thus the entanglement of the hole with the emitted radiation must keep growing as long as the physics at the horizon is `close to normal'; small corrections to each created pair cannot accumulate over a large number of emissions $N$ to lead to a decrease of $S_{ent}$.\footnote{Some authors have misunderstood the physics going into deriving the result (\ref{inequality}). For example in \cite{raju}, it was argued that one can make $S_{ent}$ decrease by making small corrections to the {\it density matrix}. But this has nothing to do with the physics at hand; one has to start with the actual process involved in Hawking emission and consider small corrections to the {\it state} obtained by evolution. Then (\ref{inequality}) shows that $S_{ent}$ cannot decrease. If we play formal games with the entries of the density matrix, then in terms of the physical state we would be making arbitrary nonlocal corrections. These nonlocalities will typically stretch over distances of order $({M\over m_p})^2 R_s\approx 10^{77} \, km$ for a solar mass hole; this is the distance over which the Hawking quanta spread during the evaporation process. By contrast the result (\ref{inequality}) allows the correction to each emitted pair to depend on all the details of what is in the hole at that emission step,  but does not bring allow arbitrary nonlocalities stretching across $\sim 10^{77}\, km$.}   By the discussion of Section  \ref{structure}, we then need to either have nonlocality, or we need to find `hair'\footnote{For further work on qubit models of Hawking radiation, see \cite{bits}.}.

\b

As we remarked above, some of the literature following AMPS suggested that AMPS had a new argument for the existence of structure at the horizon. But such is not the case.
AMPS accept Hawking's argument for structure at the horizon, and the above result of \cite{cern}. They also note that nonlocalities can lead to awkward physical observations. This leaves only the possibility of hair, and they agree that the fuzzball structure found in string theory is an example of the structure at the horizon that they invoke in their discussion. Their actual argument addresses something else: the possibility of complementarity, to which we now turn.

\subsection{Traditional complementarity}

Hawking's paradox is a serious impediment to making a quantum theory of gravity, and one might be willing to postulate new physics to bypass it. The idea of complementarity was  postulated by 't Hooft, and developed by Susskind and others \cite{complementarity}. In its initial form,   it was  a particular kind of {\it nonlocality}. Here we just summarize the main idea. One postulates that different things can be seen by different observers:

\b

(i) For the purposes of an observer outside the hole, one can imagine a stretched horizon, placed at Planck distance from the usual horizon. This stretched horizon absorbs matter falling on it, thermalizes it, and re-radiates it as low energy Hawking radiation, by a unitary process similar to one that would occur on the surface of a normal body. We call this `Picture 1'.

\b

(ii) An observer who falls into the hole does not notice the degrees of freedom of the stretched horizon;  he falls in smoothly, seeing just the  traditional black hole metric where the state around the horizon is the vacuum  $|0\rangle$. We call this `Picture 2'. 

\b

(iii) Pictures 1 and 2 can be consistent because an observer who falls in soon hits the singularity; there is not enough time for him to observe a conflict with the fact that his data has been left on the stretched horizon in a different description. 

\b

There are of course immediate objections to the set of postulates (i)-(iii)\footnote{For the precise set of postulates formulated by Susskind et al., see the first reference of~\cite{complementarity}.}. This kind of complementarity was postulated before any `hair' had been discovered at the horizon. Thus there was no known physics that reflect the information  from the stretched horizon in the manner of postulate (i). Further, the degrees of freedom on the stretched horizon are  `virtual', in the sense that they are not seen in the description of Picture 2; it is not clear what physics would lead to such degrees of freedom.

But the most serious difficulty comes from the fact that this kind of complementarity is in conflict with {\it locality}. We can study the entire process of black hole formation and evaporation using a set of `good slices' where the curvatures are always gentle everywhere; a detailed description of this slicing can be found in \cite{cern}. If we assume that normal physics holds when the curvature is low, then we get pair creation by Hawking's process, not reflection from a stretched horizon. So if we are to have complementarity, we need to find some reason why the `good slicing' through the horizon is invalid. {\it But complementarity did not provide such a reason.} Thus we are left to simply argue that whenever we have enough matter in a region to form a black hole, then some new physics takes over. Since there is no {\it local} reason to doubt normal physics along the good slices, the reason must be nonlocal, over scales of order $ r_0$, the horizon size. 

This difficulty with complementarity was of course well known\footnote{For example, in the review \cite{beyond}, it was noted: ``While this picture of complementarity appears to bypass the information problem, it is unclear how we can reconcile it with the usual idea of Hamiltonian evolution on Cauchy slices. Suppose we take the good coordinates (2.9) where a complete Cauchy surface has parts both inside and outside the horizon... If we follow Hamiltonian evolution to a later Cauchy slice, we see Hawking's pair creation and the consequent entanglement between the inside and the outside. How can we bypass the information problem that follows from this entanglement?''}.
We can finally arrive at the goal of the AMPS paper \cite{amps}. AMPS tried to make the above argument against complementarity rigorous, by using the set-up of \cite{cern}. We will summarize the AMPS argument below.  But in what follows after that,  we will argue that their discussion is misplaced in several ways, and in particular does  not address the idea of fuzzball complementarity which has been developed in \cite{plumberg,beyond}. 

\subsection{The AMPS argument}

Let us now summarize the AMPS argument. The analysis of \cite{cern} reviewed in Section \ref{sec:cern} showed that a regular horizon implies rising entanglement. Equivalently, we can say that if entanglement is to decrease, then the state at the horizon cannot be the vacuum. AMPS adapted this analysis  to suggest a crisp and elegant argument against complementarity \cite{amps}, which we summarize as follows:

\b

(A) Consider Picture 1. The radiation $\{ b_i\}$ from earlier steps of emission is near infinity. The quantum that has just been emitted, $b_{N+1}$ is outside, but close to the stretched horizon. 

\b

(B) Now consider Picture 2. We denote quanta in this picture with a prime $'$. It is assumed that everything outside the stretched horizon is identical between the two pictures:
\be
\{ b'_i\}~=~\{ b_i\}
\label{bppq}
\ee
\be
b'_{N+1}~=~b_{N+1} \,.
\label{bbp}
\ee


(C) In Picture 2 we assume that we have the vacuum at the horizon. Thus the mode across the horizon $c'_{N+1}$ is entangled with $b'_{N+1}$ in the manner assumed in Hawking's computation. Thus (\ref{inequality}) gives
 \be
  S'_{N+1} ~\gtrsim~ S'_N+\ln 2 \,.
  \ee
 (AMPS ignore the small corrections, setting $\epsilon=0$.) 
 
  \b
  
  (D) In Picture 2 we were not looking for unitarity of evaporation, since the infalling observer did not have time to measure the entanglement of emitted quanta. But by (\ref{bppq}), (\ref{bbp}) we have $S'_N=S_N$, $S'_{N+1}=S_{N+1}$. Thus we find that, in Picture 1,
  \be
  S_{N+1}\gtrsim S_N+\ln 2
  \ee
  This contradicts the fact that in Picture 1 we do want the entanglement to {\it decrease}, after the halfway point, by approximately $\ln 2$ per emitted bit.  
  
  \b
  
  This appears to be a crisp statement of the general problem that complementarity has always faced; namely, that it is not compatible with the local evolution that creates Hawking's pairs. Note that the kind of complementarity proposed by Susskind involves a very particular kind of nonlocality: 
  
  \b
  
  (i) We invoke  nonlocal effects inside the horizon to argue that we should not use the `good slicing'   that leads to Hawking's problem of growing entanglement.
  
  \b
  
   (ii) We limit this nonlocality very sharply, so that all physics outside the stretched horizon is `normal' local field theory. 
   
   \b
   
   The AMPS argument addresses this kind of complementarity, and attempts to rule it out in a rigorous way.

  \subsection{Real degrees of freedom at the horizon} 

  To see which step in the AMPS analysis needs to be altered, it is helpful to begin not with complementarity, but with the information problem. After all AMPS assume that $S_{ent}$ starts to decrease  at some point, so we should ask for the mechanism which allows Hawking's computation of rising $S_{ent}$ to be bypassed. 
  
  In string theory, we have now understood this mechanism. What we find is not a nonlocality, but a complete set of black hole `hair' -- black hole microstate solutions end in a complicated mess  just outside the place where the horizon would have formed. These `fuzzball' states radiate just like a piece of coal, so there is no information  paradox. One may think of the fuzzball surface as a stretched horizon, but note that the degrees of freedom here are {\it real}, not virtual ones that would appear only in a certain coordinate system.
  
  When fuzzball solutions were initially found, one might have argued that there were two possibilities:
  
  \b
  
  (i) That all states of the hole were fuzzballs; i.e., no state had a `traditional horizon' with the vacuum  $|0\rangle$ in its vicinity.
  
  \b
  
  (ii) That some of the states of the hole were fuzzballs, but other states would have traditional horizons. For example, as we go to more and more complicated fuzzballs, the state at the fuzzball surface could start to behave like the vacuum $|0\rangle$ for all low energy physics, in particular for the modes appearing in the Hawking process. 
  
  \b
  
  But if (ii) were true, how do we resolve Hawking's problem of growing $S_{ent}$? This is where the idea of small corrections (discussed in Section \ref{sec:cern}) came in.  Suppose that small corrections of order $\epsilon\ll 1$ could be offset by the largeness of the number of pairs $N$, so that $S_{ent}$ could start decreasing after some point instead of increasing monotonically. Then we could let the black hole states be described by the traditional Schwarzschild metric to leading order, and allow small quantum gravity effects to resolve Hawking's puzzle. Of course, in this case one would say that there was no Hawking puzzle in the first place; Hawking did only a leading order computation, and subleading effects invalidate his conclusion.
  
  But the inequality (\ref{inequality}) derived in \cite{cern} showed that small corrections do {\it not} lead to a  decrease in $S_{ent}$. We are therefore left only with the possibility (i); that is, all states of the hole are fuzzballs.
  
  With this understanding, we can now explain how complementarity {\it should} be defined.

  \subsection{The idea of fuzzball complementarity}
  
  We finally come to how complementarity should be defined in the presence of real degrees of freedom. We proceed as follows:
  
  \b
  
  (A) We have real degrees of freedom at the horizon. These degrees of freedom  radiate quanta at energies $E\sim T$ just like normal bodies do, with this radiation encoding details of the black hole state. There should {\it not} be a complementary description where the physics of these quanta is replaced by physics in the traditional black hole background. The reason is simple: the traditional hole does not exhibit the details of the black hole state, and so any radiation deduced from it it cannot reproduce the details of the state which are carried by these $E\sim T$ quanta. In particular, the relation (\ref{bbp}) should {\it not} be assumed; in general we have
  \be
  b'_{N+1}~\ne~b_{N+1} \,.
\label{bbpqq}
\ee

  \b
  
  (B) What we can hope for is the following:
  
  \b
  
  (i) We consider measurements in the frame of a lab, composed of $E\gg kT$ quanta, `falling freely from far' to the surface of the fuzzball~\cite{plumberg}.  For now, the reader can think of a lab falling freely from outside the near-horizon region; later we shall give a more precise definition of  `falling freely from far'. We describe such a process as a `hard-impact' process.

For such hard-impact processes, we conjecture that, to a good approximation, the physics is independent of which black hole microstate we have. 
  This is just like the approximation used in thermodynamics: for appropriate operators, the precise choice of microstate is irrelevant. 
  \b 
  
  (ii) We then conjecture that the physics of these hard-impact processes can be reproduced, to a good approximation, by using the traditional black hole background.
  
  \b
  
\begin{figure}[htbp]
\begin{center}
\includegraphics[scale=.62]{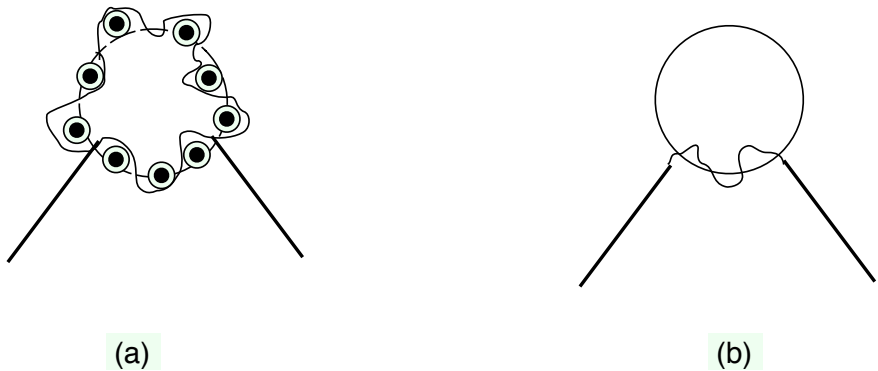}
\caption{{(a) Probing the fuzzball with hard-impact, $E\gg T$ operators causes collective excitations of the fuzzball surface. (b) The corresponding correlators are reproduced in a thermodynamic approximation by the traditional black hole geometry, where we have no fuzzball structure but we use the geometry on both sides of the horizon.}}
\label{feetwo}
\end{center}
\end{figure}

  The idea is pictured in  Fig.\;\ref{feetwo}. In Fig.\;\ref{feetwo}(a) we have the black hole microstate which ends in a fuzzball structure before the horizon is reached. Probing this fuzzball surface with operators gives correlation functions; for hard-impact processes involving $E\gg kT$ quanta, these correlators will be approximately independent of the choice of fuzzball state. In Fig.\;\ref{feetwo}(b) we have the same correlator, now measured using the traditional black hole background. Now we do not have the degrees of freedom of the fuzzball surface, but we {\it do} have the spacetime region which is the interior of the horizon.

  At first this proposal may look very strange. Do we have structure at the horizon, or don't we? If we fall onto the fuzzball surface, do we smash and burn, or don't we?  
  
  The situation is as follows:
  
  \b
  
  (a) If we write the full string theory wavefunctional of a microstate, with no approximations, we will not get the traditional hole with vacuum $|0\rangle$ at the horizon; in fact the wavefunctional is supported on configurations that end in a messy string theoretic structure before the horizon radius $r_0$ is reached.
  
  \b
  
  (b)   Now consider a hard-impact process involving $E\gg kT$ quanta. The quanta cannot penetrate the surface, since there is no `interior' region; the spacetime ends around $r\sim r_0$. One expects that the impact instead creates collective excitations of the fuzzball surface, which will spread out from the point of impact.
  
  \b
  
\begin{figure}[htbp]
\begin{center}
\includegraphics[scale=.58]{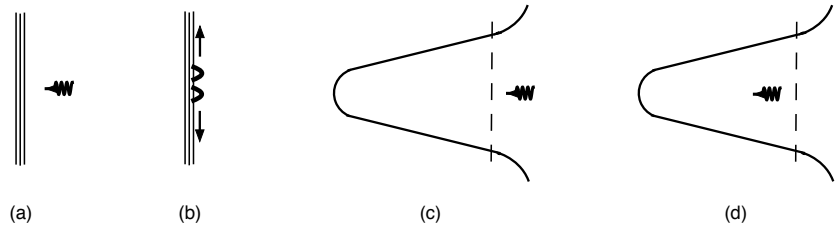}
\caption{{(a) A graviton incident on a stack of D-branes (b) The graviton `smashes' on the branes, converting its energy into a very special state of gluons (collective excitation) (c) The incident graviton in the dual AdS description (d) The graviton passes smoothly through the location where it had appeared to `smash' in the brane description.}}
\label{famps7}
\end{center}
\end{figure}

  (c) At this point one might be tempted to say that the quantum has `smashed onto the fuzzball surface', so in no sense can we say that we have fallen through. But consider the following analogy with AdS/CFT duality, introduced in \cite{mt} and pictured in Fig.\;\ref{famps7}.

 In Fig.\;\ref{famps7}(a) we depict a graviton falling onto a stack of D3 branes.  In Fig.\;\ref{famps7}(b), we see a graviton hit the branes and transfer its energy into creating collective excitations of gluons on the branes.  In Fig.\;\ref{famps7}(c) we set up the same  situation as Fig.\;\ref{famps7}(a)  in the dual gravity description.  The crucial point is seen in Fig.\;\ref{famps7}(d):  the infalling graviton does not `hit' anything at the location of the branes (indicated by the dashed line); it passes smoothly through into a new spacetime region (the AdS interior) which did not exist in the CFT description.

 So, in the AdS/CFT case, did the infalling graviton `smash and burn' on hitting the branes, or did it `feel nothing' on reaching the location of the branes? In this case the answer is clear: it is correct to say that the graviton `feels nothing' when hitting the branes. This is easy to see in the AdS description, but difficult to see in the CFT description. The key point is that excitation created on the CFT is very special: it faithfully encodes the detailed wavefunction of the infalling graviton. As discussed in \cite{mt}, a necessary (but not sufficient) condition for such faithful encoding is that the level density of states in the CFT is very high: thus a state $|\psi_E\rangle$ of the graviton with energy $E$ maps onto a state $|\psi'_E\rangle$ of the {\it same} energy $E$ on the branes, giving a unitary map
 \be
 \sum_i C_i |\psi_{E_i}\rangle ~\r~ \sum_i C_i |\psi'_{E_i}\rangle
 \ee
 which gives  a faithful injective mapping of the Hilbert space of the infalling  quantum into the  Hilbert space of the CFT. By contrast, if a graviton hits a concrete wall, it does `smash and burn', because the energy levels  available are not dense enough to exactly match the energy levels of the infalling graviton. Thus the excited state created on the concrete wall  is not a faithful map of the infalling wavefunction:
 \be
 \sum_i C_i |\psi_{E_i}\rangle ~\r~ \sum_i \t C_i |\psi'_{\t E_i}\rangle
 \ee
 and in general $\t C_i\ne C_i, ~\t E_i\ne E_i$.
 
 For the fuzzball situation, we do have a very dense set of energy levels, since the entropy $S_{bek}$ is very high. The conjecture is then that an $E\gg T$ graviton impacting hard onto the fuzzball surface creates collective excitations of a very special form, which can be given an (approximate) complementary description as free fall into the traditional hole. 
   But there is one important difference with the AdS/CFT case: the relation of Fig.\;\ref{feetwo} is an approximation valid for $E\gg T$.

\subsection{The flaw in the AMPS argument}

We can now see how the AMPS discussion heads in a wrong direction. AMPS note at one point that their discussion applies also to fuzzball complementarity. But such is not the case, since they do not address the issues relevant to such a complementarity. The following is a summary of the issues that we will encounter:

\b

(a) AMPS do not have any analogue of the condition of considering hard-impact processes of $E\gg T$ quanta. Thus they seek to get a complementary description of {\it all} processes, including those that describe the Hawking quanta with with $E\sim T$. But these quanta are supposed to carry the detailed information of the black hole microstate, and so should {\it not} be captured by any description with an `information-free horizon'. 

\b

(b) When the infalling quantum lands on the stretched horizon (the fuzzball surface), then {\it new} degrees of freedom are created, which are not entangled with the radiation at infinity. Let $N_i$ be the initial number of states of the stretched horizon; we assume that  at this point the hole is maximally entangled with the radiation at infinity. After an energy $E\gg T$ is added, the number of states is $N_f$, with
\be
{N_f\over N_i}\approx  e^{E\over T}\gg 1
\label{etp}
\ee
But only $N_i$ states of the hole are entangled with infinity; the remainder are new, unentangled degrees of freedom whose dynamics will be captured by the complementary description. Since AMPS have no analogue of the $E\gg T$ condition, they have no analogue of (\ref{etp}) either. 

   \b
   
   (c) AMPS assume that their stretched horizon stays at a fixed location $r_0$ until the infalling quantum impacts it. But this appears to be an inconsistent assumption: the stretched horizon should expand out to a larger radius {\it before} accepting the new quantum. The reason for this is that the stretched horizon at its initial radius $r_0+l_p$ encodes all the $Exp[S_{bek}]$ possible states of  mass $M$. If a quantum of energy $E$ could land on the stretched horizon  without increasing its area first, then we would have a strange situation. We would have created a new state of the stretched horizon (the one with the new quantum added) but the area would still be the old one. Thus the stretched horizon would have  {\it more} than $Exp[S_{bek}]$ states on a surface with radius $r_0+l_p$, and this contradicts the definition of the stretched horizon.  With fuzzballs, we note that the tunneling mechanism that resolves the information paradox makes the stretched horizon move out before the incoming quantum lands on it. We argue that the stretched horizon should move out by a distance
\be
s_{bubble}\sim \Big ({E\over T}\Big )^{1\over (D-2)}l_p
\ee
before the new quantum lands on it.

\b

(d) AMPS are concerned about the interaction between an infalling quantum and Hawking quanta that are escaping from the hole. We find the proper distance from the  horizon $s_\alpha$ where an incoming quantum interacts strongly with Hawking radiation quanta. We obtain
\be
s_\alpha\sim \Big({E\over T}\Big)^{1\over 2(D-2)}l_p
\ee
Thus we see that for $E\gg T$, we have 
\be
s_{bubble}\gg s_\alpha \,.
\ee
 So we can excite the new degrees of freedom on the fuzzball before we get `burnt' by interaction with Hawking radiation. These new degrees of freedom are very large in number for $E\gg T$ (Eq.\;(\ref{etp})), and are  unentangled with the radiation at infinity. The conjecture of fuzzball complementarity says that the dynamics of these new degrees of freedom is encoded in the complementary picture. 

\b

To summarize, the picture of fuzzball complementarity is different in several ways from the picture of traditional complementarity. AMPS are concerned that an infalling quantum interacts with the outgoing Hawking quanta. With fuzzballs, the Hawking quanta are just the tail end of the fuzzball surface.  The interaction of an infalling quantum with the fuzzball is `infinitely strong', in the sense that nothing can pass through the fuzzball surface -- there is no interior region to pass through {\it to}. The question then is not whether we interact with the fuzzball, it is {\it how} we interact with the fuzzball. We find that an infalling quantum  excites the new degrees of freedom (\ref{etp}) {\it before} it gets randomly scattered by the Hawking quanta. The conjecture of fuzzball complementarity then says that we  get collective oscillations of the fuzzball, which would be encoded in the complementary description in a manner similar to AdS/CFT duality.

\section{The physics of real degrees of freedom at the horizon}\label{hp}

In this section we recall the fuzzball construction which gives real degrees of freedom at the horizon. These degrees of freedom give the long-sought `hair' for the black hole, and resolve the information paradox by removing the traditional horizon from the structure of the microstate.

\subsection{The difficulty of finding `hair'}

The most direct way out of the Hawking puzzle would be to find that the state at the horizon was {\it not} the vacuum $|0\rangle$. It is this vacuum that leads to the production of entangled pairs, and if the state at the horizon was something other than $|0\rangle$, then it might be possible to have an evolution that does not produce such an entanglement. But it turned out to be remarkably difficult to find any configuration of the black hole that would  not have the vacuum $|0\rangle$ at the horizon. The geometry of the hole appears to be a standard one, determined completely to be its conserved quantum numbers, with no possibility for any deformations. This failure to find distortions of the hole is embodied in Wheeler's statement - `black holes have no hair'. 

  \begin{figure}[htbt]
\hskip .5 in\includegraphics[scale=.43]{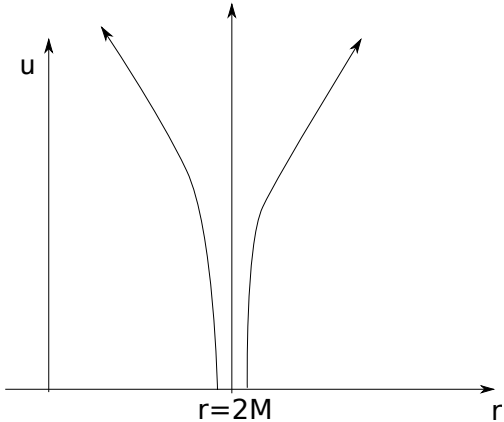}
%
%
\caption{The instability of `outgoing geodesics' at  the horizon. The horizontal axis is the radial coordinate, and the vertical axis is an Eddington-Finkelstein coordinate. A null geodesic at $r=2M$ headed radially outwards stays stuck at $r=2M$, one slightly outside escapes to infinity while one slightly inside falls to $r=0$.}
\label{fa14p}       
\end{figure}

The underlying reason for `no hair' is depicted in Fig.\;\ref{fa14p}. The figure depicts the trajectories of massless particles trying to fly radially outwards from the hole. If the particle starts outside the horizon, then it ultimately escapes to infinity. If it starts at the horizon, then it stays at the horizon forever. If it starts inside the horizon, then it is dragged to smaller $r$ and falls into the singularity. Thus the horizon is an {\it unstable} place; any particles placed there flow out  or fall in. Equivalently, we can say that the region around the horizon `stretches' due to the divergence of these trajectories, just like the stretching we get in the evolution of de Sitter space. This stretching dilutes away any matter placed at the horizon, so that after a few e-foldings the region around the horizon returns to the vacuum $|0\rangle$. The Hawking process then proceeds again, producing entangled pairs. 

AMPS try to construct a `firewall' at the horizon by placing quanta around the horizon. We note that this approach is just the same as early attempts to find `hair' for black holes. In these attempts people tried to solve the field equations for scalar, vector or gravitational fields around the horizon, looking for solutions with different angular dependences $Y_{lm}(\theta, \phi)$. If they had found such modes,  then populating them with occupation numbers $n_{lm}$ would generate a large number of states for the black hole, all different from $|0\rangle$ at the horizon. In fact, cutting off the angular quantum number $l$ when the wavelength reaches Planck scale gives roughly one mode per Planck area of the horizon, so the number of possible states $\{ n_{lm}\}$ would be order the Bekenstein entropy. But such attempts did not succeed, because there are no stable deformation modes of the horizon.  The AMPS  firewall is found to be unstable for exactly the same reason; deformations of the horizon soon return back to the vacuum.

\subsection{`Hair' in string theory: the fuzzball construction}

The required hair at the horizon was finally found in string theory, through a nonperturbative construction called the `fuzzball'. It is important to understand the nature of these fuzzball states, since any conjecture on complementarity has to start with an understanding of the structure at the horizon. 

The simplest black hole in string theory is the two-charge extremal black hole. It turned out to be possible to construct explicitly all the states of this hole, at the coupling where the black hole would be expected \cite{twocharge}. Large classes of states have now been found for more complex holes - the three-charge and four-charge holes \cite{threecharge}, and some families of states have also been constructed for the three-charge non-extremal black hole \cite{ross}. For recent work in this area, see e.g.~\cite{neckcap,superstrata1,superstrata2,deformation1,deformation2}.

\b\b

\begin{figure}[h]
\begin{center}
\includegraphics[scale=.58]{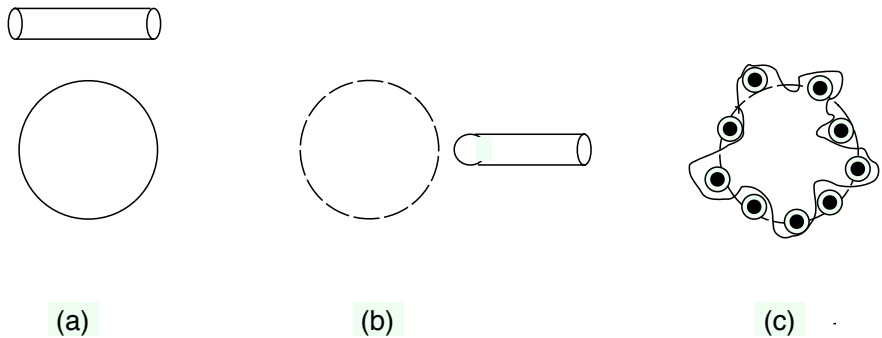}
\caption{{(a) Traditionally, it was assumed that in the black hole geometry the compact directions would appear as a trivial tensor product with the 3+1 metric. (b) In the actual microstates in string theory the compact directions pinch off to make KK monopoles/antimonopoles just outside the place where the horizon would have been. (c) The resulting solutions are `fuzzballs', which have no horizon or `interior'.}}
\label{feone}
\end{center}
\end{figure}

From these constructions we can see how the no-hair theorems are bypassed in string theory (see \cite{gibbonswarner} for a detailed discussion). A crucial role is played by the compact directions of spacetime. 
In earlier attempts to make hair, such directions were assumed to be trivially tensored with the noncompact direction (Fig.\;\ref{feone}(a)). But the states corresponding to black holes in string theory are not of this type; a compact direction pinches off before reaching the location where the horizon would have formed (Fig.\;\ref{feone}(b)). There is a  `twist' in this pinch-off, so that the overall local geometry is that of a KK monopole. At some other angular direction, the pinch-off  creates an anti-monopole, so the total monopole charge remains zero. The other objects in string theory: fluxes, strings etc. are also present, in a way that supports  these monopoles to create a full solution to string theory. The crucial point is that spacetime {\it ends} when the compact circle pinches off; there is no sense in which one can pass through the fuzzball surface into an `interior' (Fig.\;\ref{feone}(c)).

\begin{figure}[htbt]
  \begin{center}
\hskip -.5 in\includegraphics[scale=.45]{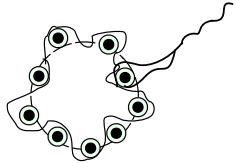}
%
%
\caption{Hawking radiation is just the tail end of the fuzzball structure that ends the geometry outside the horizon; thus there is no natural split between the degrees of freedom on the `stretched horizon' and the degrees of freedom in the Hawking radiation.}
\label{figamps1}       
\end{center}
\end{figure}

In addressing the AMPS argument below, it will also be  important to understand how Hawking radiation emerges from the fuzzball. In \cite{cm1} it was found that radiation {\it rate} from a family of nonextremal fuzzball microstates agreed exactly with the Hawking radiation rate expected for those microstates. 
This radiation does {\it not} emerge by pair creation from a vacuum $|0\rangle$ at the horizon; in fact we do not even have a horizon. What we find instead is that there are ergoregions around the KK monopoles, and the radiation occurs by the process of ergoregion emission \cite{myers}. This process creates pairs but no information paradox, since neither member of the pair is lost by falling through a horizon. The point of importance to us is that the radiation mode from the fuzzball is inseparable from the fuzzball geometry itself: the gravitational field in the ergoregion slowly evolves, creating a train of gravitational waves that become the Hawking radiation quanta as they move further out (Fig.\;\ref{figamps1}). Equivalently, we can say that the near-horizon Hawking radiation becomes very nonlinear and self-interacting as we follow it  close to the horizon, and this nonlinearity ends up creating an end to  spacetime by  pinching-off  the compact circles.

In short, the fuzzball surface and the near horizon radiation are inseparable manifestations of the same gravitational structure at the horizon, and it is not natural to break up this structure into a fuzzball surface and an emergent radiation. This fact will play an important role when we discuss the behavior objects falling onto the fuzzball. 

\subsection{Resolution of the information paradox}\label{tunnelsec}

In string theory we expect that all our states are fuzzballs with no regular horizon. Further Hawking radiation arises from the surface of the fuzzball just like radiation from a piece of coal. This resolves the information paradox.

One may now go back and ask the following.  If a shell of mass $M$ is collapsing under its own gravity, then semi-classical physics suggests that it passes smoothly through its horizon and forms the traditional black hole with horizon. How do fuzzballs alter this expectation? To see the answer, we imagine placing our entire system in a large box (of size $100 M$, say) and finding the exact energy eigenfunctions of the system $|E_i\rangle$. These state $|E_i\rangle$ are  fuzzball states which have a small `tail' in the region $r>2M$ corresponding to the radiation from the fuzzball \cite{myers,cm1}.    The collapsing shell state $|\Psi\rangle$ is a very special linear superposition of these eigenstates
\be
|\Psi\rangle=\sum_i C_i |E_i\rangle
\ee
This state evolves as
\be
|\Psi\rangle \r \sum_i C_i e^{-i E_i t}|E_i\rangle
\label{evolve}
\ee
so that it transforms to a generic superposition of fuzzball states, which then radiate from their surface in a unitary fashion. One may recast (\ref{evolve}) in the language of tunneling. There is a very small amplitude ${\cal A}\sim Exp[-S_{cl}]$ for tunneling from the shell state to a fuzzball state, where $S_{cl}$ is the action for the process. But there are a very large number of states ${\cal N}\sim Exp[S_{bek}]$ to tunnel to, and one finds that these small and large numbers are such that they may cancel \cite{tunnel, rate}. If they cancel, the wavefunction of the shell will spread over the large phase space of fuzzball states, and we can say that the large measure term (from $S_{bek}$) in the path integral destroys the semiclassical approximation.  This effect that is not present in any theory that does not have a complete set of horizonless microstates (fuzzballs).

\subsection{Behavior of the stretched horizon}

The AMPS argument assumes certain properties of the stretched horizon. In our case the stretched horizon is just the fuzzball surface, and we can check if these assumed properties are valid. In particular, AMPS assume that the stretched horizon does not respond to an infalling quantum until the quantum actually lands on it. This does not look like a consistent assumption, as we will now see:

\b

(A) First consider the stretched horizon as an abstract concept defined for the purpose of setting up complementarity. Suppose we have a black hole of mass $M$,   horizon radius $r_0$, horizon area $A$ and entropy $S_{bek}(M)$. Let the stretched horizon to be a surface just outside $r_0$. 

All the states of the hole are supposed to be given  by states of the stretched horizon. Thus the stretched horizon should have $Exp[S_{bek}]$ states. We can model these states by assuming that the stretched horizon is packed to maximum density with Planck sized cells, with each cell containing a bit in the form of a spin $s=\h$. In an `old' black hole, all these states are entangled with the radiation at infinity. Now suppose a quantum with energy $E$ lands on the stretched horizon, without creating any prior deformation of the stretched horizon. We get a new state of the stretched horizon, which is not among the $Exp[S_{bek}]$ states that we already accounted for.

\begin{figure}[ht]
\hskip 1.5 in\includegraphics[scale=.61]{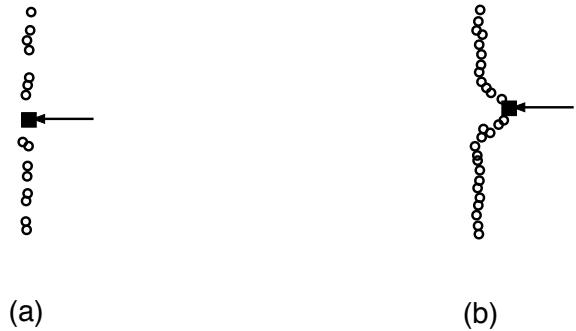}
%
%
\caption{(a) At the surface or a normal body, new quanta can be added into gaps on the surface or by pushing the existing atoms aside; this can happen because there is no constraint of having a new state on that surface. (b) On the stretched horizon, we already have the maximum number of bits that can be fitted into the area $A$; thus a new quantum can be accommodated only by first expanding the area of the stretched horizon.}
\label{famps9}       
\end{figure}

But this does not appear reasonable in the context of the picture we  have of the stretched horizon, because we expect the stretched horizon with area $A$ to have only $Exp[S_{bek}]$ possible states. To get more states, we should consider a stretched horizon with a larger area $A$. The stretched horizon is very special surface, as depicted in Fig.\;\ref{famps9}. For the surface of a normal body, depicted in Fig.\;\ref{famps9}(a), we have gaps between the atoms on the surface, so it is possible for an infalling quantum to sit in one of these gaps without any prior expansion of the surface. In Fig.\;\ref{famps9}(b) we depict the stretched horizon, which is  packed with bits to maximum density. To accept a new quantum carrying one bit of data, we would have to first expand the stretched horizon and create space for this bit. Thus we conclude that the stretched horizon should expand {\it before} it accepts a new quantum.

\b

(B) With fuzzballs, the stretched horizon (fuzzball surface) is composed of real degrees of freedom \cite{membrane}, and any dynamics of this stretched horizon should be a consequence of the normal quantum evolution (\ref{evolve}). This evolution implies that the stretched horizon `tunnels out' from its initial location to a new one when an infalling  quantum approaches.

To see this, consider the fuzzballs states corresponding to the initial mass $M$; the surfaces of these fuzzballs are located around a radius $r_0$. Now suppose a quantum of energy $E$ approaches close to the location $r_0$. We should write the state of the overall system in terms of fuzzball states with energy $M+E$; these states have their structure peaked around a new location $r_0+\delta r_0$. The `dephasing' (\ref{evolve}) converts the initial state of black hole plus shell to a linear combination of fuzzball states peaked at $r_0+\delta r_0$. One may recast this normal evolution as a `tunneling'. But note that such behavior is an essential feature in the present context because the fuzzballs of a given mass $M$ give a complete set of states at the location of the stretched horizon, and we must necessarily consider states extending to a  larger radius if we are to add energy $E$ to the system. 

Now let us return to the situation in Fig.\;\ref{famps9}(b). 
In general, if the incoming quantum deposits $E=n \, m_p$ of energy, we expect that the stretched horizon would have to expand by $n$ Planck cells before it can accept the infalling quantum. Thus the `bubble' depicted in Fig.\;\ref{famps9}(b) has area
\be
\Delta A \sim  \Big ({E\over m_p} \Big ) l_p^{D-2} \,.
\label{deltaarea}
\ee
We will assume for concreteness in what follows below that this bubble has the intrinsic geometry of a hemispherical surface protruding from the initial stretched horizon.\footnote{This appears to be a conservative assumption; in fact when two particles collide, the horizon forms in a very elongated fashion \cite{matschull}, so we may expect that the bubble stretches out even further.\label{foot:matsch}}

\section{Fuzzball complementarity}\label{fc}

We can already see that complementarity with fuzzballs will have to be  different from the way AMPS start their discussion of complementarity. Traditional complementarity relied on different reference frames to produce different effective physics; an observer staying outside the hole would see information reflected by the stretched horizon, while an observer falling in will not encounter the degrees of freedom implicit in the stretched horizon. This observer-dependence of physics was a new postulate, not something that one had encountered elsewhere in general relativity or string theory. 

But with the fuzzball construction the situation is different. The fuzzball geometry is just a regular solution to string theory; its intrinsic structure does not depend on which coordinates we use to describe it.  The geometric fact of importance is  that spacetime ends at the place where the compact circle pinches off to make monopoles/antimonopoles. In other words, we have {\it real} degrees of freedom at the stretched horizon, not virtual ones that are observer dependent. The fuzzball is really like a piece of coal, radiating quanta that carry the detailed information of its structure.

In this situation any conjecture on complementarity must take the following form:

\b

(a) The real degrees of freedom of the fuzzball have to radiate their detailed information through the energy $E\sim T$ Hawking radiation quanta. Thus it should {\it not} be possible to replace the fuzzball surface by an effective {\it vacuum} region as far as these $E\sim T$  modes are concerned. 

\b

(b) Consider a lab, composed of $E\gg T$ quanta, `falling freely from afar' to the surface of the fuzzball~\cite{plumberg}. We describe such a process as a `hard-impact' process. We can imagine that such a hard impact creates collective oscillations of the fuzzball, which are relatively insensitive to the precise choice of state of the fuzzball. It may then be possible that the Green's functions of these collective modes can be reproduced by an effective geometry that {\it does} have a smooth horizon (Fig.\;\ref{feetwo}). This effective geometry would be the complementary description. 

\b

Let us explain what we mean by $E\gg T$ quanta `falling freely from afar' into a black hole of radius $r_0$. We start at a radius $r$ with
\be
r-r_0=\beta\,  r_0, ~~~\beta >0 \,.
\label{one}
\ee
At this location, let us set up a local orthonormal frame with axes along the Schwarzschild $r, t$ directions. We require that the infalling quantum's energy $E_{local}$ (measured in this frame) be  much larger than the temperature of the local Hawking quanta (in the same frame)
\be
E_{local}\gg T_{local} \,.
\ee
We hold $\beta$ fixed, and take the mass of the hole to large values:
\be
{M\over m_p}\r\infty \,.
\label{mmp}
\ee
This is the analog of the thermodynamic approximation, where we fix the operator we wish to measure, and then take the size of the system to be large. As the ratio $M/m_p$ becomes larger and larger, we expect that our complementary description becomes more and more accurate. 

These conditions define `free fall from afar' into the black hole, and is the primary physical situation in which we seek a complementary description in terms of free fall through a horizon. 
The condition $\beta>0$  implies that we can start falling from infinity, or from a location $r=3 r_0$, or even $r=1.1r_0$. By contrast, suppose we  start at rest from a location that is given to be a fixed distance from the horizon: e.g., $r-r_0=100 l_p$ and then fall in. This is a {\it fine-tuned} process, which corresponds to gently lowering the quantum to a point near the horizon, and then letting it fall in. We do not require a complementary description of such fine-tuned processes, since such processes are the analog of measuring a fixed number of atoms in a gas: the thermodynamic limit (\ref{mmp}) does not improve the fluctuations in the measurement.

\subsection{The statement of fuzzball complementarity}

In \cite{plumberg} the  complementarity depicted in Fig.\;\ref{feetwo} was formulated.  This approach is based on earlier work of Israel \cite{israel}, Maldacena \cite{eternal} and Van Raamsdonk \cite{raamsdonk}, where the black hole spacetime was written as an entangled sum of states (see \cite{beyond} for a full discussion).

\begin{figure}[htbp]
\begin{center}
\includegraphics[scale=.65]{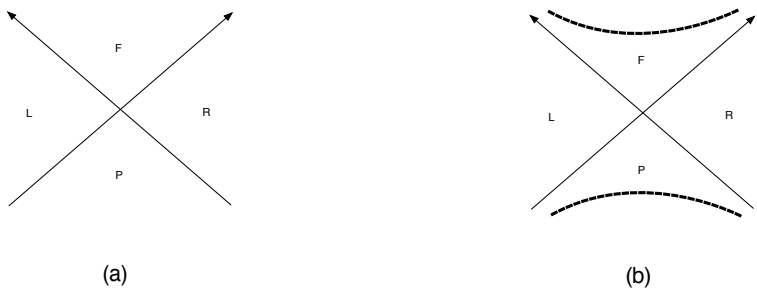}
\caption{{(a) Minkowski space and its Rindler quadrants (Right, Left, Forward and Past). (b) The Penrose diagram of the extended Schwarzschild hole. The region near the intersection of horizons is similar in the two cases.}}
\label{fn5}
\end{center}
\end{figure}

The essential idea can be seen from Fig.\;\ref{fn5}. Minkowski space can be decomposed into Rindler quadrants. If we consider a scalar field $\phi$, then the Minkowski vacuum state $|0\rangle_M$ can be written as an entangled sum of states  in the left and right Rindler quadrants:
  \be
|0\rangle_M=C\sum_i e^{-{E_i\over 2}}|E_i\rangle_L|E_i\rangle_R, ~~~~~~~C=\Big (\sum_i e^{-E_i}\Big )^{-\h}
\label{split}
\ee
For a scalar field $\phi$, the states $|E_i\rangle_L, |E_i\rangle_R$ are known explicitly in terms of Bessel functions. One may ask:  what is the corresponding split for the gravitational field? In particular, 
the central region of the eternal black hole is similar to Minkowski space (Fig.\;\ref{fn5}(b)), so we may expect a similar decomposition of the eternal black hole geometry
\be
|g\rangle_{eternal}=C\sum_k e^{-{E_k\over 2T}}|g_k\rangle_L\otimes |g_k\rangle_R, ~~~C=\Big (\sum_i e^{-{E_i\over T}}\Big )^{-\h}
\label{qwfourt}
\ee
But what are the `Rindler' states $ |g_k\rangle$? Most of the entropy for a given mass $M$  comes from states that are {\it black holes}. But if the black hole solution is the traditional one with a smooth horizon, then how do we distinguish different Rindler states $|g_k\rangle$? This puzzle goes away when we recognize  that black hole microstates are fuzzballs, which have nontrivial structure were the horizon would have been. In fact the fuzzballs are the natural candidates  for the Rindler states of gravity, since they end (by the pinch-off of a compact circle) in their own Rindler quadrant, without leaking past the horizon. 

We can recover Minkowski space as the central region of the black hole in the limit $M\r\infty$. The fuzzball states appear when we decompose the Minkowski vacuum state in gravity into left (L) and right (R) pieces across a boundary surface.  In particular we find that Rindler and de Sitter entropies can be obtained as the entanglement entropy of the fuzzball states appearing in such decompositions \cite{beyond}.

We can now arrive at complementarity by putting together two observations:

\b

(A) Suppose we compute the expectation value of an operator $\hat O_R$ which is localized in the region outside the horizon of the black hole, but where we assume that the state corresponds to the full eternal black hole.   Noting the decomposition (\ref{split}), we find
\bea
{}_{eternal}\langle 0|\hat O_R|0\rangle_{eternal}&=&C^2\sum_{i,j}e^{-{E_i\over 2T}}e^{-{E_j\over 2T}}{}_L\langle g_i|g_j\rangle_L \, {}_R\langle g_i|\hat O_R|g_j\rangle_R\nn
&=&C^2\sum_i e^{-{E_i\over T}}{}_R\langle g_i|\hat O_R|g_i\rangle_R
\label{qwe1}
\eea
Thus the expectation value in the eternal black hole geometry  is given by  a thermal average over fuzzball states.

\b

(B) A given black hole is in {\it one} fuzzball state. But for a generic fuzzball state, and for appropriate operators $\hat O_R$, we can approximate the expectation value by the ensemble average over all fuzzballs
\be
{}_R\langle g_k|\hat O_R|g_k\rangle_R\approx {1\over \sum_l e^{-{E_l\over T}}}\sum_i e^{-{E_i\over T}}{}_R\langle g_i|\hat O_R|g_i\rangle_R \,.
\label{qwe2}
\ee
Putting together (\ref{qwe1}) and (\ref{qwe2}) we get\footnote{Recently an approach to complementarity was developed in \cite{malsuss}, which also used the  idea of \cite{eternal} about entangled giving the eternal black hole. It should be noted that our proposal of fuzzball complementarity is quite different: in particular \cite{malsuss} do not attempt to get complementarity through an approximation of a microstate by its canonical ensemble. For us, the approximation sign in (\ref{qwe2}) is vital, because the real degrees of freedom on at the horizon have a double duty: they have to carry details of the microstate for $E\sim T$ (so there is no complementarity for such modes), and they have to provide a complementary description of $E\gg T$ infall. Ref.\;\cite{malsuss}, on the other hand, seeks to get a complementary description for all modes.}
\be
{}_R\langle g_k|\hat O_R|g_k\rangle_R\approx {}_{eternal}\langle 0|\hat O_R|0\rangle_{eternal} \,.
\label{qwe3}
\ee
 This is the statement of fuzzball complementarity. That this is a kind of complementarity can be seen as follows:
 
 \b
 
 \begin{figure}[h]
\begin{center}
\includegraphics[scale=.75]{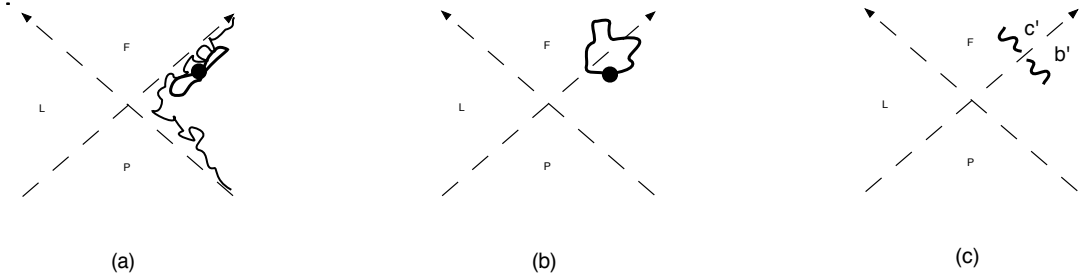}
 \caption{{(a) Computing a 1-point function in a generic fuzzball geometry; the operator excites the complicated mess of degrees of freedom on the fuzzball surface (Picture 1). (b) For appropriate operators, the same correlator can be obtained to a good approximation by using the traditional black hole geometry; now there are no degrees of freedom at the horizon but the paths in the path integral can smoothly cross across the horizon into a new auxiliary spacetime region - the black hole interior (Picture 2). (c) In Picture 2, the state is such that we do have modes $b', c'$ entangled as expected in the vacuum.}}
\label{famps2}
\end{center}
\end{figure}

 (i)  On the LHS we compute with a gravitational state which has no horizon; spacetime ends in a messy quantum set of degrees of freedom before the horizon is reached.   We illustrate this in Fig.\;\ref{famps2}(a), where we illustrate the computation of $\langle\hat O_R\rangle$ schematically by a path integral. The operator $\hat O_R$ excites the complicates degrees of freedom of the fuzzball. 
 \b
 
 (ii) On the RHS we compute with a spacetime which has a smooth horizon where the gravitational field is in the {\it vacuum} state $|0\rangle$. In computing $\langle\hat O_R\rangle$ in Fig.\;\ref{famps2}(b) the paths in the path integral smoothly cross the horizon back and forth, noticing nothing special at the horizon. If we break the the vacuum $|0\rangle$ into modes $b', c'$ across the horizon, then these modes are indeed correctly entangled to make the vacuum state (Fig.\;\ref{famps2}(c)).

 \b
 
 But the entanglement of $b',c'$ noted in (ii) is exactly what AMPS had argued should not be found. Thus we should now go back and see if one or more of their assumptions were inappropriate. 
 As we will discuss in more detail below, the crucial point will be that they do not invoke any approximation of the kind in (\ref{qwe2}) which forces the complementarity to be accurate only for `appropriate' operators. 
 The notion of appropriate operators here is the similar to the notion of appropriate operators in statistical mechanics, where only for appropriate operators can a generic microstate be replaced by a canonical ensemble average. 
 
 With fuzzballs, we see that Hawking quanta carry the information of the microstate, and so detailed measurements of Hawking quanta cannot be in the class of appropriate operators. Since these quanta have energy $E\sim T$, where $T$ is the temperature of the hole, we can say that the complementary description should somehow involve excitations with $E\gg T$.\footnote{Note that the value of $T$ depends on the location and the reference frame; the temperature is high near the horizon in the Schwarzschild frame, and the Hawking quanta at this location have a correspondingly high energy.} 
 Fuzzball complementarity focuses on hard-impact processes involving $E\gg T$ quanta as the processes of physical interest, and conjectures that they are described by the traditional black hole.\footnote{A faulty understanding of fuzzball complementarity appeared in \cite{raju}; it is important to clear this misconception. This paper claims that if an observer measures one $E\sim T$ quantum, then (according to fuzzball complementarity) he should have a full identification of the microstate. This is of course incorrect (and fuzzball complementarity does not say this). 
Fuzzball complementarity states that hard-impact processes involving $E\gg T$ quanta are the
processes of interest for the idea of complementarity, and conjectures that they are described by the traditional black hole. 
 Of course there are other simple processes involving $E\sim T$ quanta for which the predictions of a typical microstate and the classical black hole agree, e.g.~due to general statistical reasoning like that used by Page~\cite{page}.}

Note that the decomposition of the Minkowski vacuum $|0\rangle_M$ into Rindler states is an operation done on {\it one} time slice. The same is true of the decomposition of $|0\rangle_{eternal}$. Correspondingly, the operator $\hat O_R$ in (\ref{qwe1}) should be  an operator defined on {\it one} time slice. In this situation we get the approximation (\ref{qwe3}). But one can also consider processes where the relevant operators cannot be made to lie on one time slice. This situation, for the black hole case, will occupy us in what follows.  We will comment on the analogous situation for Minkowski space in the discussion  at the end of the paper.

\subsection{The failure of the relation $b=b'$}\label{difference}

\begin{figure}[htbp]
\begin{center}
\hskip -1 in \includegraphics[scale=.65]{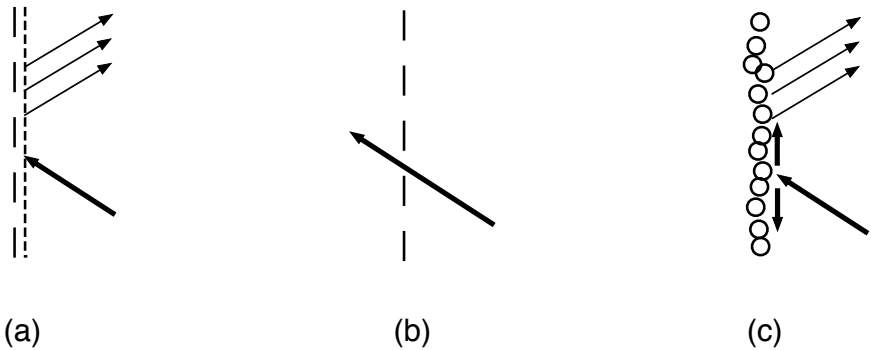}
 \caption{{(a) Traditional complementarity: we imagine that infalling matter is absorbed and re-radiated from virtual degrees of freedom at a stretched horizon; these degrees of freedom are virtual because we have found no `hair' for the hole. (b) In the complementary picture, an infalling observer fails to see anything at the location of the stretched horizon. (c) The situation with fuzzballs: real degrees of freedom end the spacetime outside the horizon. These degrees of freedom must  play two roles: (i) radiate Hawking quanta at $E\sim T$ and (ii) account for any complementarity through collective oscillations generated by hard impacts of $E\gg T$ quanta.}}
\label{famps3}
\end{center}
\end{figure}

We can now pinpoint where fuzzball complementarity differs from the kind of complementarity assumed by AMPS (Fig.\;\ref{famps3}). In fuzzball complementarity the Hawing modes $b$ outside the horizon in Picture 1 do not agree with the corresponding modes $b'$ in Picture 2:
\be
b\ne b' \,.
\ee
 These Hawking modes are the $E\sim T$ quanta that are supposed to differentiate microstates, and so they are {\it not} going to be reproduced by the complementary description. By contrast, traditional complementarity does not provide any explanation for how Hawking's puzzle will be resolved. AMPS therefore start from the assumption
 \be
 b=b' \,.
 \ee
 Because of this assumption, they miss the possibility of fuzzball complementarity.
 
\subsection{The significance of $E\gg T$}\label{nfnisec}

Let us now see in more detail how the condition $E\gg T$ will be relevant in bypassing the argument of AMPS. Among all operators  with  $E\gg T$  are those that excite collective modes of the fuzzball. Such collective modes, in turn, can be excited when the fuzzball is impacted by heavy objects falling in from far outside the hole. Thus we are led to ask the following question. Suppose a quantum with $E\gg T$ falls freely from afar onto the fuzzball surface. In this situation can we conjecture a  dynamics where we have a  `complementary' description? 

When we impact a star, we do not expect complementarity, so what is special about impacting the surface of a fuzzball? The key issue is the fact that the density of states of the fuzzball, given through $S_{bek}$, grows extremely rapidly with energy.  Thus the temperature
\be
T={dE\over dS}={dM\over dS}
\ee
is very small when the  mass $M$ of the hole is large. 

Suppose we start with a black hole of mass $M$. This hole has  $N_i=Exp[S_{bek}[M]]$ possible fuzzball states $|E_k\rangle$. Now suppose we throw in a quantum with energy $E\gg T$. The total energy is now $M+E$, and the number of possible states is $N_f=Exp[S_{bek}[M+E]]$. We have
\be
{Exp[S_{bek}[M+E]]}= Exp[S_{bek}[M] + \Delta S]\approx Exp[S_{bek}[M]+{E\over T}]=Exp[S_{bek}[M]]\,  e^{E\over T}
\ee
where in the second step we have used the thermodynamic relation $dE=TdS$.
Thus we find
\be
{N_f\over N_i}={Exp[S_{bek}[M+E]]\over Exp[S_{bek}[M]]}\approx  e^{E\over T}
\label{nfniq}
\ee
For $E\gg T$ we get
\be
{N_f\over N_i}\gg 1
\label{nfni}
\ee 
This means that when we impact the fuzzball at high energy, most of the phase space allowed consists of {\it new} states that were not accessible before the impact.

Let us consider this observation in the context of the AMPS argument. AMPS wait until the black hole has evaporated past its half-way point, so that its states are maximally entangled with the emitted radiation $R$:
\be
|\Psi\rangle_{initial}={1\over \sqrt{N_i}}\sum_{i=1}^{N_i} |E_i\rangle \otimes |R_i\rangle
\ee
 But (\ref{nfni}) shows that when the maximally entangled hole is impacted by a quantum with $E\gg T$, then most of the allowed states of the hole will be new states. These new states are {\it not} entangled with the radiation at infinity. This is the case because only $N_i$ states of the hole are entangled with the radiation,  but the total number of possible states is $N_f\gg N_i$. The complementarity conjecture  pertains to the dynamics of these unentangled $N_f-N_i\approx N_f$ states. We can conjecture an evolution of these states which would correspond to collective modes of the fuzzball, and look for a complementary description of this evolution.\footnote{This creation of new states is the content of the equation $ \sum_i C_i |F_i\rangle\r\sum_j C'_j |F'_j\rangle$ noted in \cite{mt}. The relevance of new states is also noted in \cite{ver}.}

\subsection{The analogy between fuzzball complementarity and AdS/CFT}

We do not, of course, know the full dynamics of fuzzball states under an impact with $E\gg T$, so we cannot give the complementarity map in explicit form. What we do here is conjecture a possible dynamics of fuzzballs to show how the AMPS argument fails to address fuzzball complementarity. 

There are two main features of fuzzball complementarity: (i) the approximation $E\gg T$ and (ii) The notion that Green's functions arising from collective modes of the fuzzball can be obtained by using the traditional black hole geometry as an auxiliary spacetime. While (i) is peculiar to the context of black holes emitting Hawking radiation,  (ii) is the analogous to the idea of AdS/CFT duality. We use this duality as an analogy, introduced in \cite{mt}, to help set up our conjecture on collective dynamics of fuzzballs. The following comments should help clarify this analogy:

\b

\begin{figure}[ht]
\begin{center}
\hskip -.5 in 
\includegraphics[scale=.58]{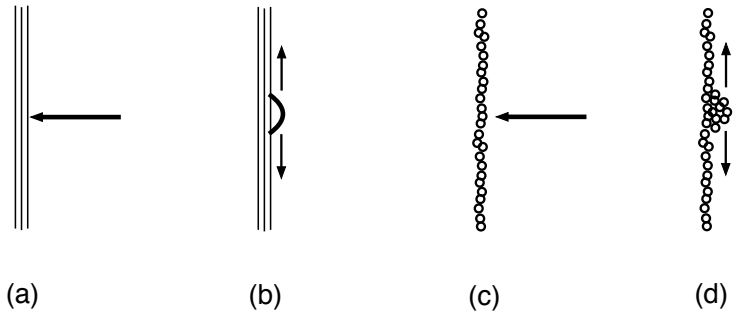}
 \caption{{(a) A quantum is incident on a stack of D-branes; (b) The quantum `smashes' on the D-branes, but its energy gets converted into  ripples that spread out on the branes; (c) The analogous situation with the fuzzball surface: a graviton is incident on the monopoles etc.~making the fuzzball; (d) The graviton `smashes' onto the fuzzball surface, converting its energy to ripples that spread along the fuzzball surface.}}
\label{famps4}
\end{center}
\end{figure}

(A) In Fig.\;\ref{famps4}(a) we depict a quantum falling onto a stack of D-branes. The impact creates a disturbance of  gluons on the branes; this disturbance spreads  away from the impact point in ripples moving out at the speed of light (Fig.\;\ref{famps4}(b)).

In Fig.\;\ref{famps4}(c) we depict a high energy quantum impacting the fuzzball surface. We imagine that the impact creates a large number of new monopoles/antimonopoles at the impact point. This excitation then relaxes, with the monopole structure spreading out from the impact point in circular ripples. The dynamics of this ripple  is the collective dynamics of the fuzzball. 

\b

\begin{figure}[ht]
\begin{center}
  \includegraphics[scale=.65]{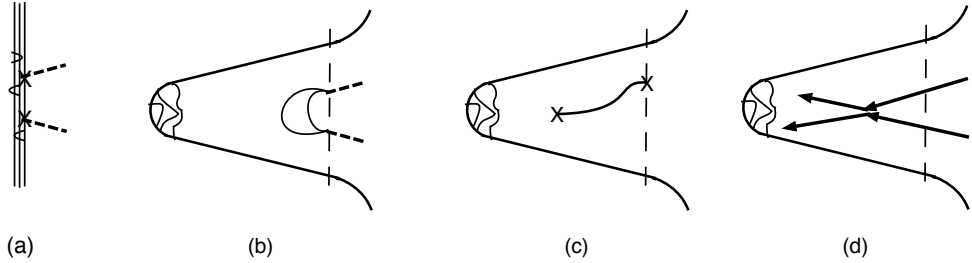}
 \caption{{(a) A 2-point function in the CFT: there are some low energy excitations present, but the correlator of interest is not sensitive to them; (b) The dual gravity picture: the low energy excitations are localized deep in the AdS, they do not interfere with the paths contributing to the correlator; (c) One can ask about boundary-bulk propagators in the gravity picture, but these must be re-expressed in terms of boundary-boundary propagators before comparisons can be made with the CFT; (d) A typical process of interaction in the bulk: while this is encoded in the boundary CFT, it is very difficult to extract this encoding from  correlators  in the CFT.  }}
\label{famps8}
\end{center}
\end{figure}

(B) In Fig.\;\ref{famps8} we describe how quantities are measured in the AdS/CFT correspondence:

\b

(i)  In Fig.\;\ref{famps8}(a) and Fig.\;\ref{famps8}(b) we consider the computation of a 2-point correlator in the CFT and AdS pictures, with a view to understanding the significance of  the requirement $E\gg T$ in the fuzzball case.  We consider a situation where some low energy excitations are already present on the D-branes; these are described by gravity quanta lying deep in the AdS region. We now insert two local operators to measure a `high energy' correlator. The inserted operators create high energy excitations, which do not interact strongly with the low energy excitations already present on the branes. In the gravity description we see this lack of interaction by the fact that the paths contributing to the correlator lie near the AdS boundary, away from the location of the low energy quanta. Thus we get a kind of universality: if we go to high energies, we can expect to generate collective excitations that are insensitive to the precise starting state of the system.

The computation of Fig.\;\ref{famps8}(a) corresponds to the computation in Fig.\;\ref{feetwo}(a), and that in Fig.\;\ref{famps8}(b) corresponds to Fig.\;\ref{feetwo}(b). The fact that the precise choice of initial state did not matter in Fig.\;\ref{famps8} corresponds to the fact that the correlator in Fig.\;\ref{feetwo}(a) is independent of the precise choice of fuzzball. 

\b

(ii)  One may ask: what happens if we fall onto the stack of D-branes?  In the CFT, we `smash', creating a large number of excitations that encode the data of the infalling object. In the dual gravity description we pass smoothly into the AdS interior. In the latter case, we might now want to ask about experiments conducted by the infalling observer after he falls through into the AdS region. These questions involve a boundary to bulk propagator of the kind depicted in Fig.\;\ref{famps8}(c), or bulk-to-bulk propagators. But the bulk points are not points in the CFT, so what do such questions mean? 

The answer is simple: the operator in the bulk can be expressed as a  (complicated) combination of operators in the boundary, and thus correlators like Fig.\;\ref{famps8}(c) can be rewritten as linear combination of correlators Fig.\;\ref{famps8}(b) . A similar situation holds for fuzzball complementarity. The only true questions are those that can be measured at or outside fuzzball surface, as in Fig.\;\ref{feetwo}(a). The complementary description Fig.\;\ref{feetwo}(b) obtains approximate answer to these questions by using an auxiliary spacetime that has  a smooth horizon leading to an interior region. One may try to start backwards and ask questions about operators in the interior region in Fig.\;\ref{feetwo}(b), but we would have to relate these `interior' operators to operators that  lie on the fuzzball surface before we can use fuzzball complementarity to relate the two pictures. 

\b

(iii) We pursue this further in Fig.\;\ref{famps8}(d), where we depict a simple question in the bulk. We have two high energy quanta falling into the AdS, and we  can ask if they collide before reaching the bottom of AdS.  In the black hole case the corresponding question relates to one  inside the horizon: `Do two particles in a lab collide before the lab hits the singularity?'. Again, we note that the question in Fig.\;\ref{famps8}(d) can be asked in the dual CFT, but obtaining the answer will be very complicated. Each infalling particle creates a set of expanding ripples on the CFT surface, and we are asking a delicate question about the intersection of these ripples. In particular, the products of the collision in Fig.\;\ref{famps8}(d) will fall into the bottom of AdS, and mix with the low energy quanta present there. Thus  in the CFT we will not know if the particles collided or not until we examine all  the data encoded in the CFT state. Analogously, in the black hole case  the infalling quanta of Fig.\;\ref{feetwo}(a) will generate ripples on the fuzzball surface, but whether the quanta collided or not in Fig.\;\ref{feetwo}(b) before hitting the singularity will be something that is delicately encoded in all the data on the fuzzball surface.

\b

With these remarks, we proceed to examine experiments with fuzzballs in more detail.

\section{Bypassing the AMPS argument}

The AMPS argument requires that the hole be maximally entangled with the radiation $R$ at infinity, but we have seen in (\ref{nfni}) that upon impact by a quantum with $E\gg T$, most of the accessible states are {\it not} entangled with $R$. Have we therefore completely bypassed the AMPS argument?

The answer is yes, provided we check one thing: these new unentangled states can be accessed before the infalling quantum is `scattered' by interaction with Hawking quanta. 
In this section we will give the computations which show that the scattering is indeed too small to prevent complementarity. Before we start the computation, let us review our goal.

AMPS are concerned that an infalling quantum interacts with outgoing Hawking radiation. But we have seen that with fuzzball complementarity, this is not the correct question. The Hawking quanta are the `tail end' of the nonlinear fuzzball structure that ends outside the horizon location. Thus there is no way that we can ever avoid interaction with the Hawking radiation -- we cannot pass through the fuzzball surface to an interior, since there {\it is} no interior. The correct question is not {\it whether} we interact with the Hawing radiation; it is {\it how} we interact with this radiation. 

Thus consider a quantum with energy $E$ at infinity, falling towards the hole. There are two possibilities:

\b

(i) The probability of significant interaction with Hawking quanta far from the stretched horizon is low. The principal interaction occurs near the stretched horizon, where the new, unentangled degrees of freedom (\ref{nfni}) are excited. The state created with these new degrees of freedom in insensitive to the precise initial state of the fuzzball. The situation is then like that in AdS/CFT duality, in the following sense. The infalling quantum has interacted (strongly) with the nonlinear part of the Hawking radiation - i.e., the structure near the fuzzball surface. But this interaction is similar to the one where an infalling quantum smashes onto a collection of D-branes. Then the excitations generated on the fuzzball surface can have a complementary description where it falls smoothly through the horizon. 

\b

(ii) The probability of significant  interaction with Hawking quanta far from the stretched horizon is {\it not} low. In this the infalling quantum will be scattered into a new state, and this new state will depend on the precise state of the radiation emerging from the black hole microstate. In this situation we will still get new, unentangled degrees of freedom (\ref{nfni}) when the quantum reaches the fuzzball surface, but since the state of the quantum has already been altered by scattering, the excitations on the fuzzball surface cannot be given a complementary description where `nothing happens' as we fall through a horizon.

\b

The physical processes of interest are hard-impact processes involving $E\gg T$ quanta `falling freely from far', as described in (\ref{one})--(\ref{mmp}).
In this section we will see that for such processes we get (i) rather than (ii). Where appropriate, for concreteness we shall consider the example of a Schwarzschild black hole in $D$ spacetime dimensions. As before, $r_0$ shall denote the horizon radius. 

In the rest of this section, we proceed in the following steps: 

\b

(a) In Section \ref{timeqq} we begin with a warm-up exercise. AMPS envision measuring the Hawking quanta $b$. We show that a detector `falling freely from far', as specified in the condition (\ref{one}), cannot perform such a measurement because there is too little proper time left along its infall trajectory to make the measurement.  This is a `warm-up exercise' in the sense that it illustrates the constraints that arise when we use the $E\gg T$ condition, but it is not an important step in what follows; we are really interested in scattering off $b$ quanta, rather than the ability to perform detailed measurements on them. 

\b

(b) In Section \ref{cross} we compute the cross section for scattering the infalling quantum off a Hawking radiation quantum. We find that the scattering can be split into two classes: a large impact parameter part where we get gentle spin-preserving deflections, and a hard scattering part where the state of the infalling quantum is altered. 

\b

(c) In Section \ref{compute} we compute the distance $s_\alpha$ from the horizon where hard scattering becomes significant. We find that $s_\alpha<s_{bubble}$, where $s_{bubble}$ is the distance from the horizon where the new degrees of freedom (\ref{nfni}) are expected to be accessed. 

\b

(d) In Section \ref{scatter} we show that the `gentle deflection' part of the scattering does not distort the incoming quantum significantly by the time it reaches the fuzzball surface.

\b

Putting all these computations together, we conclude that interactions with Hawking quanta do not preclude the possibility of fuzzball complementarity.

\subsection{The difficulty of `measuring' a $b$ quantum}\label{timeqq}

AMPS imagine Gedanken experiments that measure the state of a $b$ quantum near the horizon. As a warm up exercise for the computations in the next section, it will be useful to note that such a measurement cannot be performed by observers that fall freely towards the hole from afar. 

Suppose we try to construct an apparatus that falls in towards the black hole and `measures' a $b$ quantum before crossing the horizon. As an example, we may hope to detect quanta of energy $\sim 1\,  TeV$, which have a  wavelength    $\lambda\sim 10^{16} l_p$.  We should make our infalling detector from quanta that interact with the radiation quanta. Further, we should be able to switch the detector on and off appropriately; i.e., switch it on when we reach the location where the $b$ quantum is expected, and switch it off  before the detector falls through the horizon. 

We will now observe that under very general assumptions, it is not possible to construct such an apparatus. More precisely, we will find that the interaction with the emerging radiation is low unless we get very close to the fuzzball surface, and once we have fallen that close to the fuzzball surface there is not enough proper time for the detector to respond to the measurement. 

In detail, we proceed as follows:

\b

(A) We assume that our detector is composed of quanta that, individually, have energies lower than Planck energy
\be
E\lesssim m_p \,.
\label{four}
\ee
This relation is satisfied, of course, by all the detectors that we have in our laboratory. (Violating this condition will force us to consider string states; such states will have a size that grows with $E$, and an interaction cross section that needs to be computed using details of string theory.)\footnote{If we use purely ingoing massless quanta to make the detector, then the proper time along the detector trajectory will be zero, and the detector cannot respond to any interaction. Nevertheless, when the detector falls close to the horizon, the quanta composing it would be moving with a speed close to the speed of light in the Schwarzschild frame. We will deduce interaction cross sections from the cross sections for massless gravitons, since all non-stringy states in string theory have interactions that are not too different from the interaction of gravitons. For example, we can imagine that our particles have a mass $m$ because they carry a momentum around a compact direction of length $L\gtrsim l_p$, and the interactions of such quanta are obtained from the interactions of 10-d gravitons in string theory.} 

With the detector made of quanta satisfying (\ref{four}), we find on general grounds that the typical frequencies that can be detected will lie in the range $\omega\lesssim m_p$, and the time $t_{switching}$ in which the detector can be switched on and off  must  satisfy
\be
t_{switching}\gtrsim l_p \,.
\label{five}
\ee

\b

(B) We will allow the detector to fall in freely from afar; i.e., we let it fall from a position satisfying (\ref{one}).

\b

(C) We assume that our theory of gravity is string theory, where the low energy quanta are in the graviton multiplet,  and thus have all their interactions determined by a universal coupling encoded in the Newton's constant $G$. The interaction between gravitons grows with the energy of the gravitons. As the infalling quantum comes closer to the horizon, the probability of interaction with a radiation quantum increases. This increase happens for two reasons: the infalling quantum becomes more energetic (as measured in a local static frame) and the radiation quanta encountered also have a higher energy.

Let us measure radial positions in terms of the proper distance $s$ from the horizon,  measured along a constant $t$ slice. Let the infalling quantum be a graviton with energy $E$ at infinity, and let $T\sim {1\over r_0}$ be the Hawking temperature. In Appendix \ref{interact} we show that the location where the probability of interaction $P$ with a radiation quantum becomes order unity is given by
\be
s_{interact}\sim  ({E\over T})^{1\over 4} l_p \,.
\ee
If the infalling quantum had started at a finite radius $\bar r$ instead of at infinity, then the value of $s$ where $P\sim 1$ would be even smaller. Thus we write
\be
s_{interact}\lesssim  ({E\over T})^{1\over 4} l_p \,.
\label{twoq}
\ee
Noting that $T\sim {1\over r_0}$, and using (\ref{four}) we can rewrite this as
\be
s_{interact}\lesssim  ({r_0\over l_p})^{1\over 4} l_p \,.
\label{two}
\ee

\b

(D) Now we consider the proper time $\tau_{available}$ available on the trajectory of the infalling quantum, between the time it is at $s= s_{interact}$ and the time where it crosses the horizon at $s=0$. In Appendix \ref{time} we consider a particle which starts at rest at $r=\bar r$ and falls to the location which is at proper distance $s$ from the horizon. We assume that the starting value $\bar r$ satisfies (\ref{one}). Then we find that 
\be
\tau_{available}\sim ({s_{interact}\over r_0})\, s_{interact}\ll s_{interact} \,.
\label{three}
\ee

\b

(E) Putting together (\ref{two}) and (\ref{three}) we find
\be
\tau_{available}
~\lesssim~  ({l_p\over r_0})^\h l_p 
~\ll~ l_p \,.
\label{insuff}
\ee
We had noted in Eq.\;(\ref{five}) that we cannot switch on and off our measuring device in a time less than $l_p$. By contrast, here we see that we are required to perform this switching in a time that is smaller than $l_p$ by a parametrically small factor $({l_p\over r_0})^\h$. We conclude that we cannot make a detector that will perform the postulated measurement of the quantum $b$.

\subsection{The probability of interaction with a Hawking radiation quantum}\label{cross}

We now come to the real question that is relevant to the AMPS discussion: in what manner does  an infalling quantum scatter off Hawking radiation quanta?

As in the above discussion, we measure radial positions in terms of the proper distance $s$ from the horizon,  measured along a constant $t$ slice. Let the infalling quantum be a graviton with energy $E$ at infinity, and let $T\sim {1\over r_0}$ be the Hawking temperature. On dimensional grounds one may write for the interaction cross section
\be
\sigma\sim G^2  (\omega\omega')^{D-2\over 2}
\label{none}
\ee
where $\omega,\omega'$ are the energies of the infalling and emerging gravitons measured in the local orthonormal frame. We have (see Appendix \ref{interact})
\be
 \omega\sim E (g_{tt})^{-\h}\sim \Big ( {E\over T}\Big ){1\over s}, ~~~~~~\omega'\sim {1\over s} \,.
 \label{naatwo}
\ee
 Of course the total cross section may diverge because of infrared effects leading to near-forward scattering, but let us use (\ref{none}) as an estimate of the probability of significant scattering. It is shown in Appendix \ref{interact} that the probability of interaction $P$ between the infalling graviton and  a Hawking quantum becomes order unity when we reach $s$ of order
\be
s_{interact}\sim  ({E\over T})^{1\over 4} l_p \,.
\ee
We can go to a center of mass frame where the two quanta have equal energy; this requires boosting along the radial direction by a boost factor
\be
\gamma\sim ({E\over T})^\h \,.
\ee
The energies of the two colliding quanta in this center of mass frame are
\be
E_{cm}\sim \Big({E\over T}\Big)^\h{1\over s} \,.
\ee
For $s\sim s_{interact}$, we have
\be
E_{cm}\sim \Big({E\over T}\Big)^{1\over 4}m_p
\ee
where $m_p$ is the Planck mass. For $E\gg T$, we find
\be
E_{cm}\gg m_p \,.
\ee
Thus the interaction is at a transplanckian energy, and we need some understanding of gravitational collisions at such energies. We use the analysis of \cite{banks} where it was argued that we can break up the interaction into three classes: 

\b

(i) When the impact parameter is sufficiently small, the collision leads to  a black hole. Let the center of mass energy of the interaction  be $E_c$, and let a black hole with mass $E_c$ have radius $R_s$. If the impact parameter is $d\lesssim R_s$, then the interaction leads to the formation of a black hole. 

(ii) There is an intermediate range of impact parameters 
\be
R_s\lesssim d\lesssim \alpha R_s
\ee
 where a black hole may not form, but the interaction is strong and dependent on the details of the gravity theory. Here $\alpha$ is a constant factor, which does not scale with any parameters of the problem.

(iii) For $d\gtrsim \alpha R_s$ we have weak gravitational scattering given by classical physics. The interacting quanta deflect by a small angle, following geodesics in a weakly curved metric.

\b

We now combine (i) and (ii) into a single category, thus obtaining the following two domains to examine:

\b

(a) The domain with impact parameters
\be
d\lesssim \alpha R_s
\ee
where the gravitational interaction is strong. We will find that the probability of such interactions is parametrically small in our context. 

\b

(b) The domain 
\be
d\gtrsim \alpha R_s
\ee
in which the deflection is parametrically small. The issue of importance for us is that in this domain the motion of the interacting quanta is like the deflection of light by the sun: the quantum moves along a geodesic which is deflecting through a small angle. In such a process the spin of the quantum does not change, and we will see that  the infalling graviton  reaches    the surface of the fuzzball to generate collective oscillations. We have a similar situation in strong interaction physics when  a high energy quantum passes through a nucleus: the spin of the high energy quantum does not flip, and we generate collective excitations of the nucleus. 

\subsection{The domain $d\lesssim \alpha R_s$}\label{compute}

In Appendix \ref{bhsigma} we find the probability $P_{\alpha}$ that a quantum that starts at infinity with energy $E$ interacts with a Hawking radiation quantum with impact parameter  $d\lesssim \alpha R_s$. From Eq.\;(\ref{nbh}) we see that the probability $P_{\alpha}$ becomes order 1 when $s$ is of order
\be
s_{\alpha}\sim \Big({E\over T}\Big)^{1\over 2(D-2)}\alpha^{D-3\over D-2}l_p\sim \Big({E\over T}\Big)^{1\over 2(D-2)}l_p
\label{nnbhp}
\ee
where in the last step we noted that $\alpha$ is a constant which does not scale with any parameters of the problem. Note that
\be
{s_\alpha\over s_{interact}}\sim \Big ( {E\over T}\Big )^{-{D-4)\over 4(D-2)}}
\ee
so for $D>4$ the value of $s$ where the quanta interact strongly is less than the value given by naive dimensional analysis.

We must now compare $s_\alpha$ with the distance scale where we expect the tunneling effects discussed in Section (\ref{tunnelsec}) to take place. Since we are throwing in an energy $E$, the entropy of the hole will increase by $\Delta S \sim {E\over T}$, which corresponds to an area increase (using (\ref{deltaarea}))
\be
\Delta A_E\sim \Big ({E\over T}\Big ) l_p^{D-2} \,.
\label{deltaae}
\ee
We have assumed  that this area deformation starts with a hemispherical bubble of area $A_E$ around the point where the infalling quantum would impact the surface. (That is, we assume that if we go to the locally flat coordinates around the horizon, the intrinsic geometry of the bubble is that of a hemisphere. See also the comment in Footnote \ref{foot:matsch}.) Such a hemisphere would have radius
\be
s_{bubble}\sim A_E^{1\over D-2}\sim \Big ({E\over T}\Big )^{1\over (D-2)}l_p \,.
\label{bubble}
\ee
We note that
\be
{s_{bubble}\over s_\alpha}\sim \Big ({E\over T}\Big )^{1\over 2(D-2)} \,.
\ee
Then for sufficiently large $E/T$ we find
\be
{s_{bubble}\over s_\alpha}\gg 1
\ee
so we expect to excite the new degrees of freedom on the fuzzball surface much before we interact strongly with a Hawking radiation quantum. Finally, we note that (\ref{deltaae}) corresponds to creating a very large number of new degrees of freedom given by (\ref{nfniq}). The encoding of information into the fuzzball surface starts when the phase space volume of newly excited degrees of freedom becomes comparable to the phase space volume of already existing degrees of freedom. In \cite{emission} a method was developed, in  related computation, for estimating the point where 
\be
{N_{new}\over N_i}\gtrsim e\approx 2.7 \,.
\ee
It was found that this point corresponds to a distance much larger than the analog of $s_{bubble}$. Thus one may expect that in our present case the encoding of data into the fuzzball degrees of freedom starts well before the location $s_\alpha$.\footnote{In Appendix \ref{timetunnel} we estimate the time along the infalling trajectory for this tunneling to occur.}

\subsection{The domain $d\gtrsim \alpha R_s$}\label{scatter}

In this domain of impact parameters, we have gentle deflection of the incoming graviton, with no change of its polarization. Such deflections are similar to the gravitational deflections encountered by a photon as it travels to us from a distant star. The light from the star reaches us at a different point than it would if there were no deflections, but the image of the star is still reproduced faithfully since the relative separation of photons and their spin do not alter significantly under such deflections. We will now check that the deflections suffered by out infalling graviton are sufficiently small so that they do not alter the image of the infalling object on the fuzzball surface.

We will now return to using the cross section $\sigma$ suggested on dimensional grounds (which is larger than the cross section $\sigma_\alpha$ for hard interactions).  The question that we must address is depicted in Fig.\;\ref{famps6}:

\begin{figure}[htbp]
\begin{center}
 \includegraphics[scale=.73]{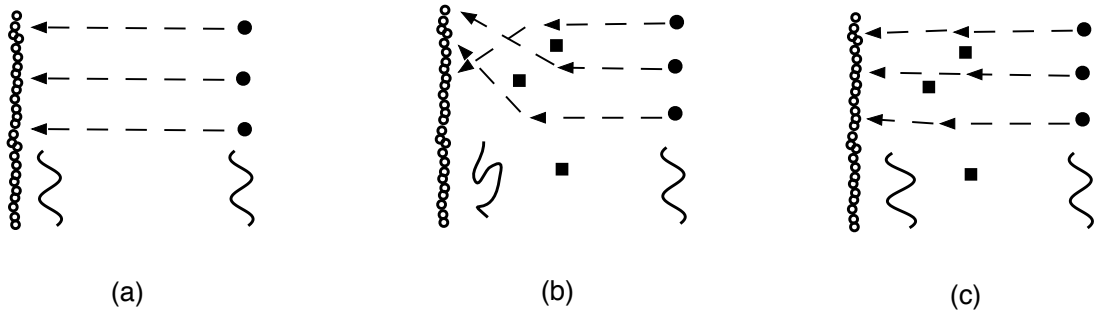}
 \caption{{(a) A body made of three atoms (solid dots) falls towards the fuzzball surface. Alternatively, we may consider the wavefunction of a quantum (wavy line). (b) The situation when the interactions with the Hawking radiation quanta are significant: the atoms get scrambled, and the wavefunctions gets severely distorted, all before reaching the fuzzball surface. (c) The situation where the deflections by the Hawking quanta are small; the infalling object retails its character till it reaches the new degrees of freedom available on the fuzzball surface.}}
\label{famps6}
\end{center}
\end{figure}

\b

(i) In Fig.\;\ref{famps6}(a) we depict the infalling object as an array of particles stretching along a direction $y$ transverse to the radial direction.  These particles have a  transverse separation $d_y$. If nothing impedes the infall of the object, then the particles reach the fuzzball surface with separation $y$, and create an excitation that leaves a faithful imprint of the object.

\b

(ii) In Fig.\;\ref{famps6}(b) we depict a situation where the particles of the infalling object hit the $b$ quanta near the horizon and scatter. Each particle suffers some deflection $\Delta y$, so that it reaches the fuzzball surface with a different value of $y$ from the one that it started with. In this figure we have depicted the situation where
\be
{\Delta y\over d_y}\gtrsim 1 \,.
\ee
In this situation when the particles reach the fuzzball surface and access new states, the data of the infalling object has become distorted. Any complementary description provided by excitation of the fuzzball surface 
will not be a faithful representation of the initial state of the object.

\b

(iii) In Fig.\;\ref{famps6}(c) we again have an interaction between the particles in the object and the $b$ quanta, but we depict a situation where 
\be
{\Delta y\over d_y}\ll 1 \,.
\label{dydy}
\ee
In this case the  object reaches the fuzzball surface without significant distortion, and the $N_f-N_i$ new states accessed at the fuzzball surface can capture its details and provide a complementary description.

\b

In Appendix \ref{deflection} we show that the interaction of infalling objects with Hawking quanta $b$ gives us the situation (iii), not the situation (ii). More precisely, we use the estimate of interaction cross section in Appendix \ref{interact} to recall that the probability of interaction of the infalling quanta with a $b$ quantum is small until we reach a distance $s\sim  ({E\over T})^{1\over 4} l_p$ from the horizon (see Eq.~\ref{aaone}). Here $E\gg T$ is the energy of the infalling particle at infinity. At the point of interaction the infalling quantum is highly blueshifted, so the collision process takes place in a frame that is boosted by  $\gamma\sim \sqrt{E\over T}\gg 1$ compared to the center-of-mass frame for the collision. One then finds for the transverse deflection
\be
\Delta y ~\lesssim~ l_p({E\over T})^{-{1\over 4}}~\ll~ l_p \,.
\label{deltay}
  \ee
By contrast, the wavelength of the particles making the object and the separation between particles are all assumed (Eq.\; \ref{four}) to be gives by a scale $d_y\gtrsim l_p$. Thus for sufficiently large $E/T$ we find $\Delta y\ll d_y$, which is the desired relation (\ref{dydy}).\footnote{We can also consider the scattering of the infalling quantum off {\it charged} Hawking quanta. In string theory we can obtain such charged quanta from gravitons carrying momentum along a compact directions. But such quanta have a mass $m\ne 0$, and are thus localized within a (fixed) distance $s\sim m^{-1}$ of the horizon. By contrast the distance $s_{bubble}$ rises with $E/T$, so for large $E/T$ we will not be scattered by charged Hawking quanta. (We thank Don Marolf for raising the issue of charged Hawking quanta.)}

The `gentle deflections' that we get from the large impact parameter scattering have an  analog in the AdS/CFT case. Consider the infall  in Fig.\;\ref{famps8}(d). The quanta deep in the AdS lead to small deformations at the `neck' of the geometry where the quantum enters the AdS; these deformations are different for different choices of the quanta placed deep in the AdS. The effect of this neck deformation on the infall is only slight; the spin and internal structure of the infalling state is not altered by such deformations, and we have smooth infall in spite of these deformations.

\subsection{Summary of the above computations}

Let us summarize the computations above that show how the AMPS argument is bypassed:

\b

(a) The idea of fuzzball complementarity is that an infalling object  does not avoid interaction with the structure at the horizon (fuzzball surface); rather, the object `smashes' onto this structure in the same way that it would smash into a collection of D-branes. The excitations on the fuzzball surface can then have a `complementary' description involving smooth infall, just like the interaction with D-branes could be replaced by smooth infall into AdS. There is one difference however from the AdS case: for the black hole case we expect the complementary description to become a good approximation only for hard-impact processes involving quanta with energies much higher than the Hawking temperature: $E\gg T$.  

\b

(b) AMPS use the fact that the states of the hole are maximally entangled with the radiation at infinity after the half-way point of evaporation. But in Eq.\;(\ref{nfni}) we see that when a quantum with $E\gg T$ impacts such a black hole, then most of the accessible states of the hole are {\it not} entangled with the radiation at infinity. These newly accessible states are the analogue of the gluons that we create in AdS/CFT when the graviton hits the D-branes. It is these newly accessed, unentangled  degrees of freedom whose dynamics will be captured by the complementary description.

\b

(c) There is no fundamental distinction between the fuzzball surface and the Hawking quanta in the near-horizon region: the radiation in this near horizon region is just the tail end of the full nonlinear structure at the fuzzball surface. Equivalently, we can say that the Hawking radiation becomes more and more strongly self-interacting as we closer to the fuzzball surface, and this self-interaction leads to the pinching-off of compact directions to form KK monopole-like structure at the horizon (where for generic states, it is understood that we are extrapolating to Planck scale degrees of freedom). Thus the fuzzball construction provides a complete set of `hair' for the hole with the following structure: the spacetime ends in this KK monopole structure before a horizon is reached, and the gravitational tail of this structure is the Hawking radiation. 

\b

(d) When the infalling quantum comes close to the fuzzball surface, its energy $E$ leads to the creation of new KK monopole-antimonopole pairs; these are the new unentangled degrees of freedom (\ref{nfni}). We can then expect a complementary description of these new states. There is however a  possible problem:  the infalling quantum might first scatter off the Hawking quanta that are far from the fuzzball surface; i.e., it hits the `tail-end' of the fuzzball structure in such a way that its state changes $\psi_i\r \psi_f$. In that case we would have a difficulty: even if the quantum finally reaches close enough to the fuzzball surface to create new degrees of freedom (\ref{nfni}), the state it creates there would reflect $\psi_f$ rather than $\psi_i$, so we would not have a complementary description where `nothing happened' as we fell through the horizon.

\b 

(e) To address the above concern, we computed the interaction of an infalling quantum with the Hawking quanta. We find that hard scattering happens at a distance $s_\alpha$ from the horizon. But the new degrees of freedom are accessed at a distance $s_{bubble}$ from the horizon, where $s_{bubble}\gg s_\alpha$ for $E\gg T$. This removes the concern noted in (d), and so fuzzball complementarity is not ruled out by an AMPS type argument.\footnote{AMPS had noted that one may consider that the stretched horizon moves out before the infalling quantum hits. But they dismissed this possibility, saying that in such a situation they can just consider interaction with Hawking quanta further out~\cite{amps}. But we see that such is not the case: the interaction with Hawking quanta becomes significant at a given place since Hawking radiation has a very special (low) temperature, and the stretched horizon automatically moves out to $s_{bubble}\gg s_\alpha$ for $E\gg T$ by the same tunneling process that leads to a resolution of the information paradox~\cite{tunnel,rate}.} (There is also a `long-distance' component to the scattering, but this does not alter the state $\psi_i$ of the infalling quantum.)

\b

(f) So we bypass the AMPS argument, and are back to the picture of fuzzball complementarity described in \cite{mt}: the infalling quantum (with $E\gg T$) hits the fuzzball and excites collective excitations on its surface; the dynamics of these collective excitations is captured in a complementary description of smooth infall.

\b

Let us make a comment regarding the validity of effective field theory when we have structure at the horizon as described in Section \ref{structure}. If one asks about all possible low-energy processes, effective field theory is not valid when we have such structure at the horizon.
However, the main claim of fuzzball complementarity is that hard-impact processes involving high-energy quanta have an approximate dual description in terms of an effective field theory. The above results show that the AMPS argument does not rule out such a complementarity.

\b\b

\subsection{Comment on the Rindler decomposition of Minkowski space}

Take a very large black hole with its traditional geometry, and consider a point just outside the horizon. The local geometry approximates that of Rindler space, and extending the Rindler metric past the horizons gives Minkowski space. This suggests that the idea of fuzzball complementarity developed above should also apply to the Rindler decomposition of Minkowski space. We now comment on this.

As is well known, if we take any time slice of Minkowski space, then around any point on such a slice  we can make a decomposition of the Minkowski vacuum into a left and a right Rindler wedge, as in Eq.~(\ref{split}). The conjecture of \cite{plumberg} is that for such a decomposition, we find fuzzball states $|E_i\rangle_R, |E_i\rangle_L$ on the right and left sides. So far, this applies on any given time slice, and is exact.

Let us now investigate whether such a decomposition may also be used to give some description of the process of a quantum passing through a Rindler horizon.
Consider a quantum of mass $m>0$ moving with some energy $E$, starting in the right Rindler wedge. For simplicity, let us assume that the particle is not accelerating relative to the Minkowski background.
Let the particle be incident on the Rindler horizon.

Let us consider the description of this process provided by a decomposition of the Minkowski vacuum into left and right Rindler fuzzballs. Of course, there are many such decompositions, depending on the choice of Rindler frame relative to the rest frame of the quantum.
We seek a decomposition which has the following property: when the quantum impacts the surface of the right Rindler fuzzball $|E_i\rangle_R$, the subsequent evolution of the excitation of this surface gives a good encoding of the information in the quantum. 
Of course, in Minkowski space, the quantum passes through the Rindler horizon in a perfectly smooth way. 
The question here is whether the Rindler fuzzball decomposition can reproduce this physics in any Rindler frame, to a good approximation.

From what we have learnt above, the encoding of information becomes better the higher the value of the energy $E'$ with which the quantum impacts the fuzzball, where here $E'$ is measured in the rest frame of the fuzzball. In the case of the black hole, the energy $E'$ at impact depended on the energy of the quantum at infinity and also on the black hole mass $M$. Thus the accuracy of encoding became better for larger $M$.

We now observe that in the case of Minkowski space, by choosing the Rindler frame appropriately we can make the accuracy as high as we want. 
For a black hole there is a particular rest frame in which the fuzzball surface is at rest; this is the Schwarzschild frame where the metric appears time independent. However when choosing a Rindler frame, firstly we may choose the origin, and secondly, having fixed the origin, we can consider boosts on the $T, X$ coordinates of Minkowski space. Using these freedoms, we can arrange that when the quantum impacts the fuzzball surface it does so with an energy $E'$ which can be chosen to be as large as we wish. In particular, for any given starting distance from the origin, this can be achieved using the boost freedom. See Appendix \ref{rindlercalcs} for a demonstration of this.  With such a choice of frame for the Rindler decomposition, we get an accuracy for our `fuzzball complementarity' which is as good as we wish.

\section{Summary}

The AMPS argument and the discussion around it have provided a good opportunity for us to understand the new progress on black holes in the context of the longstanding puzzles in the area. We have seen that the pieces of the puzzle have been available for some time, but the way they fit together is very different from what most people had imagined. Let us review this overall picture.

\b

(A) AdS/CFT duality~\cite{adscft} is a beautiful relation among degrees of freedom in string theory. Many people thought that this duality conjecture solved the information paradox. As explained in \cite{cern}, this is not the case; if anything, evidence for the correspondence {\it sharpens} the paradox. The reason is simple. For energies below the black hole threshold we get agreements between the CFT and gravity correlators, so we have more reason than ever to believe that the black hole radiates  like a normal body, with information emerging in the Hawking radiation. But above the black hole threshold people just wrote down the AdS-Schwarzschild metric, and for this metric the Hawking computation gives monotonically  growing entanglement $S_{ent}$ so information {\it cannot} emerge in the radiation. This conflict is the paradox.\footnote{See \cite{beyond} for a more detailed discussion.}

One might hope that Hawking's argument was faulty, because small corrections to his leading order computation could alter his conclusion -- these corrections  might cause $S_{ent}$ to start reducing at some point. But the inequality (\ref{inequality}) derived in \cite{cern} showed that this hope is false: there is no way to get information in Hawking radiation unless the corrections at the horizon are order {\it unity}. 

\b

(B)  The information paradox is solved in string theory by actually finding the construction of black hole `hair': the fuzzballs are very nontrivial solutions of string theory, carrying the quantum numbers of the black hole, but ending in a quantum mess before reaching the horizon radius  $r_0$. Some people thought that fuzzballs only gave some microstates of the hole, and that the other microstates would have a traditional horizon to a first approximation. But the inequality (\ref{inequality}) shows that in this case we cannot get information in the Hawking radiation. Other people thought that as we move from simple fuzzball solutions to more complicated ones, the fuzzball solution approaches the traditional hole with vacuum $|0\rangle$ at the horizon. This is false as well: the evolution of Hawking modes does {\it not} tend towards the evolution in vacuum as we go to more complicated fuzzballs. (If it did tend towards vacuum evolution, we would again have the information problem, by (\ref{inequality}).) In short, all states of the hole must be fuzzballs where the Hawking modes do not evolve the way they do in vacuum.

 \b
 
 (C) Finding the fuzzball structure of microstates solves the information paradox; the fuzzball radiates from its surface like a normal body. But now we can ask a different question: is there any significance to the traditional black hole metric? There are two aspects to the answer. First, if we look at the {\it Euclidean} Schwarzschild metric, then it provides a saddle point for the  path integral over all gravitational solutions. Thus individual states in the Lorentzian section are fuzzballs, but the thermal partition function over all fuzzballs can be expressed through a Euclidean path integral. The Gibbons-Hawking analysis then suggests that the Euclidean Schwarzschild solution is a saddle point for the path integral; this saddle point is spherically symmetric, while the individual Lorentzian microstates are not\footnote{Subleading corrections to the path integral can be found by expanding around the saddle point; thus  the metrics used in  computations like \cite{sen} should be thought of as Euclidean rather than Lorentzian.}.
 
 But what about the Lorentzian section of the traditional black hole metric? Following \cite{israel,eternal,raamsdonk} we developed the notion of fuzzball complementarity where for hard-impact processes with $E\gg T$ we can replace the complicated fuzzball surface by the eternal black hole metric. The physics is analogous to AdS/CFT: impacting the fuzzball surface hard creates ripples on this surface, just like hitting a collections of branes creates ripples on the branes. The complementary description is infall into the eternal hole in the former case, and infall into AdS in the latter. The only difference is that we need the condition $E\gg T$ for complementarity in the black hole case, since we should have {\it no} complementarity for the modes that carry the information of the microstate.
 
 Thus AdS/CFT does not resolve the information paradox, but it provides the crucial understanding of the `infall problem'. These are opposite problems in a sense. Solving the information problem requires a mechanism for emission that distinguishes microstates, and so requires the construction of `hair'. The infall problem asks if in some approximation all microstates behave the {\it same}, so that we may reproduce their dynamics by the traditional metric which has an `information free horizon'.  
  
 \b
 
 (D) Traditional complementarity was conjectured at a time when we did not have a construction of hair to solve the information paradox; thus it was formulated as a set of rules that could {\it evade} the information problem at the expense of a certain kind of nonlocal physics. It is therefore not surprising that the discussion around the AMPS paper leads right back to Hawking's paradox, stated in H' in Section \ref{intro}.
 
Today we understand the quantum structure of black hole microstates, and the resolution of the information paradox. Bringing in this perspective, we can see where the AMPS discussion goes wrong.  In particular, AMPS are concerned that an infalling quantum will interact with the Hawking quanta. But as we have seen, such interaction is bound to occur when the microstates are fuzzballs, since the Hawking radiation is just the tail end of the fuzzball, and one cannot penetrate the fuzzball surface because there is no `interior'. The correct question is not {\it whether} we interact with the radiation; rather it is {\it how} we interact with this radiation. If hard-impacts of $E\gg T$  quanta  excite collective modes on the fuzzball surface, then the dynamics of these modes is what will be described by the complementary description.
 
The issue now is to check that an infalling quantum can excite collective  excitations (using the new degrees of freedom on the fuzzball surface) before it suffers random scattering by Hawking quanta further away from the horizon. Our observation $s_{bubble}\gg s_\alpha$ (for $E\gg T$) indicates that we can in fact get the required collective oscillations; thus we bypass the AMPS argument. 

We noted that even if one did not use any details of fuzzball dynamics, the properties assumed by AMPS for the stretched horizon appear to be inconsistent. The stretched horizon of area $A$ encodes all $Exp[S_{bek}]$ states of the hole; indeed AMPS assume a maximal entanglement with all such states. If a new quantum with energy $E$ could land on this stretched horizon without prior deformation of the stretched horizon, then we would have more than $Exp[S_{bek}]$ states with area $A$, and this does not appear natural with any definition of the stretched horizon. 

To conclude, in this paper we have found that with fuzzball dynamics, the tunneling process that solves the information paradox makes the fuzzball surface move out by tunneling in a way that would allow complementarity to work.

\section*{Acknowledgements}

This work was supported in part by DOE grant DE-FG02-91ER-40690. We would like to thank Finn Larsen, Emil Martinec and Herman Verlinde for helpful comments and discussions. We are particularly grateful to Don Marolf and Joe Polchinski for discussing their work over detailed correspondence.

\begin{appendix}

\section[Appendices]{Interaction cross-section for gravitons} \label{interact}

We consider a graviton starting at infinity with energy $E$, falling radially in the Schwarzschild metric in $D$ spacetime dimensions
\be
ds^2=-(1-{r_0^{D-3}\over r^{D-3}})dt^2+{dr^2\over 1-{r_0^{D-3}\over r^{D-3}}}+r^2 d\Omega_{D-2}^2
\label{apptwo}
\ee
It will be useful to define $s$ to be the radial distance measured from the horizon $r=r_0$ along a constant $t$ slice
\be
s=\int_{r'=r_0}^r {dr'\over (1-{r_0^{D-3}\over r'^{D-3}})^\h}\approx \sqrt{4r_0\over D-3}\sqrt{r-r_0}
\label{apps}
\ee
where the second relation is for points close to the horizon
\be
{r-r_0\over r_0}\ll 1
\label{appone}
\ee

We wish to estimate the interaction cross section $\sigma$ between this infalling graviton and gravitons that are emerging as Hawing radiation. We proceed in the following steps:

\b

(A)  Let the interaction occur at a distance $\sim s$ from the horizon; we will assume (\ref{appone}) in what follows. It will be useful to refer all computations  to an orthonormal frame  $(\hat t, \hat r)$ at this location  with axis along the $t, r$ directions. By dimensional analysis
\be
\sigma\sim G^2  (\omega\omega')^{D-2\over 2}
\ee
where $\omega,\omega'$ are the energies of the infalling and emerging gravitons measured in the local orthonormal frame. We have
\be
 \omega\sim E (g_{tt})^{-\h}\sim E {r_0\over s}
 \label{aatwo}
\ee

\b

(B) The Hawking radiation at distance $s$ from the horizon is described as follows. The typical quantum has wavelength $\lambda\sim s$, and the separation between quanta is also by distances of order $\lambda\sim s$. Consider the region where distances from the horizon lie in the interval $(s, 2s)$. An infalling geodesic will cross $\sim 1$ quanta in this interval, and the encountered quantum will have
\be
\omega'\sim {1\over s}
\label{appthir}
\ee
The interaction cross section is then
\be
\sigma\sim G^2 (\omega\omega')^{D-2\over 2}\sim G^2 ({Er_0\over s^2})^{D-2\over 2}
\ee
The Hawking quantum occupied a traverse area 
\be
a\sim \lambda^{D-2}\sim s^{D-2}
\ee
Then the probability of interaction with the encountered Hawking quantum is
\be
P\sim {\sigma\over s^{D-2}}\sim G^2 ({Er_0\over s^2})^{D-2\over 2} {1\over s^{D-2}}\sim ({l_p\over s})^{2D-4} ({E\over T})^{D-2\over 2}
\label{apptw}
\ee
where we have noted that $G$ is related to the Planck length $l_p$ by $G\sim l_p^{D-2}$, and  the Hawking temperature is
\be
T\sim {1\over r_0}
\ee
We note in passing that if the infalling quantum also had energy of order the Hawking temperature, $E\sim T$, then the interaction probability $P$ becomes order unity at $s\sim l_p$; i.e., at Planck distance from the horizon. 

\b

(C) To find the total probability of interaction with the Hawking radiation quanta, we should take a value  $\bar s\sim r_0$, and consider intervals of $s$ of the kind $(2\bar s, \bar s)$, $({\bar s\over 2}, \bar s)$, $({\bar s\over 4}, {\bar s\over 2})$ etc, and add the probabilities of interaction in each interval. But due to the rapid rise of $P$ with decreasing $s$, the sum is dominated by the interval that is closest to the horizon. Thus if we fall in from infinity to a distance $s$ from the horizon, the total probability of interaction is
\be
P\sim ({l_p\over s})^{2D-4} ({E\over T})^{D-2\over 2}
\ee
We find that $P\sim 1$ when
\be
s\sim  ({E\over T})^{1\over 4} l_p \,.
\label{aaone}
\ee

\b

(D) Let us also note what would happen if we took an interaction cross section that did not grow with the energy $\omega$. Let the interaction cross section between the infalling quantum and a graviton be $\sigma_0$. Then the probability of interaction would be
\be
P\sim {\sigma_0\over s^{D-2}}
\ee
so we get $P\sim 1$ when
\be
s\sim \sigma_0^{1\over D-2}
\ee
This is a fixed distance from the horizon in Planck units. On the other hand, the energy of the infalling quantum at this location is
\be
E_{local}\sim {Er_0\over s}\sim E{r_0\over \sigma_0^{1\over D-2}}
\ee
Thus any effects that grow with $E_{local}$ will dominate over the interaction with Hawking quanta  in the limit $r_0\r \infty$.

\refstepcounter{section}
\section*{\thesection \quad Proper time along an infalling geodesic} \label{time}

We consider with a particle falling radially in the metric (\ref{apptwo}). Let the particle start from rest at $r=\bar r$, where it has proper velocity
\be
U^\mu=(U^t(\bar r), 0)
\ee
(We will write only the $t,r$ components since only these will be nonzero.) From $U^\mu U_\mu=-1$ we find $U^t(\bar r)=\sqrt{-{1\over g_{tt}(\bar r)}}$. The quantity $U_t=g_{tt} U^t$ is conserved along the motion; at $r=\bar r$ we find the value
\be
U_t=-\sqrt{-g_{tt}(\bar r)}
\label{appthree}
\ee
At position $r$, we have
\be
U^\mu=(U^t, U^r), ~~~
g_{tt}(U^t)^2+g_{rr}(U^r)^2=-1
\ee
Using (\ref{appthree}), and noting that in the metric (\ref{apptwo}) we have $g_{rr}=-{1\over g_{tt}}$,  we get
\be
(U^r)^2=-g_{tt}(\bar r)+g_{tt}\approx -g_{tt}(\bar r)
\label{appseven}
\ee
where we have assumed that $\bar r-r_0\sim r_0$, and the value of $r$ where $U^r$ is now being computed satisfies (\ref{appone}). We have
\be
g_{rr}\approx ({2r_0\over (D-3)})^2{1\over s^2}
\label{appsix}
\ee
Thus the radial proper velocity in the local orthonormal frame is
\be
{ds\over d\tau}=U^rg_{rr}^\h=-({2r_0\over (D-3)}){1\over s}(-g_{tt}(\bar r))^\h
\ee
The time to cross the horizon from a position $s$ is then
\be
\tau={d-2\over 2r_0}{1\over (-g_{tt}(\bar r))^\h}\int_0^s ds' s'={D-3\over 2r_0}{1\over (-g_{tt}(\bar r))^\h}s^2
\label{appeight}
\ee
To summarize, suppose we start at rest at $\bar r$ with $\bar r-r_0\sim r_0$, and fall to a point close to the horizon (proper distance from horizon $s\ll r_0$).  Along this freely falling trajectory, there will be a very short proper time 
\be
\tau\sim ({s\over r_0})\, s\ll s
\ee
left for the evolution from the position $s$ to the horizon.

\newpage

\refstepcounter{section}
\section*{\thesection \quad Cross section for black hole formation} \label{bhsigma}

In this Appendix we compute the cross section for the formation of a black hole upon collision of an infalling quantum  with a Hawking radiation quantum.

Let the infalling quantum  start with an energy $E$ at infinity. At a distance $s$ from the horizon, the energy of this quantum in the local orthonormal frame is
 (Eq.\;(\ref{aatwo})) 
\be
\omega\sim E{r_0\over s}\sim {E\over T} {1\over s}
\ee
and the Hawking quantum $b$ has energy  (Eq.\;\ref{appthir}))
\be
\omega'\sim {1\over s}
\ee
We can go to a center of mass frame where these two quanta have equal energy; this requires boosting along the radial direction by a boost factor
\be
\gamma\sim ({E\over T})^\h
\ee
The energies of the two colliding quanta in this center of mass frame are
\be
E_{cm}\sim ({E\over T})^\h{1\over s}
\ee
A black hole with this energy will have a radius
\be
R_s\sim (l_p^{D-2}E_{cm})^{1\over D-3}\sim \Big(l_p^{D-2}({E\over T})^\h{1\over s}\Big)^{1\over D-3}
\ee
The cross section for black hole formation is given on noting that the impact parameter should be $d\sim R_s$. We then find
\be
\sigma_{bh}\sim \Big(l_p^{D-2}({E\over T})^\h{1\over s}\Big)^{D-2\over D-3}
\ee
Following the argument leading to (\ref{apptw}), we have for the probability of such a black hole forming interaction 
\be
P_{bh}\sim {\sigma_{bh}\over s^{D-2}}\sim \Big(l_p^{D-2}({E\over T})^\h{1\over s}\Big)^{D-2\over D-3}{1\over s^{D-2}}
\ee
We find $P_{bh}\sim 1$ when
\be
s\sim \Big({E\over T}\Big)^{1\over 2(D-2)}l_p
\label{nbh}
\ee
If we allow a somewhat larger range of impact parameters $d\lesssim \alpha R_s$, then we find a cross section
\be
\sigma_\alpha\sim \alpha^{D-2}\sigma_{bh}
\ee
The probability of interaction is
\be
P_{\alpha}\sim {\sigma_{\alpha}\over s^{D-2}}\sim \Big(l_p^{D-2}({E\over T})^\h{1\over s}\Big)^{D-2\over D-3}\alpha^{D-2}{1\over s^{D-2}}
\ee
We get $P_\alpha\sim 1$ when
\be
s\sim \Big({E\over T}\Big)^{1\over 2(D-2)}\alpha^{D-3\over D-2}l_p
\label{nbhp}
\ee

\refstepcounter{section}
\section*{\thesection \quad Time available for tunneling} \label{timetunnel}

From (\ref{bubble}) we see that tunneling effects should start when the infalling quantum is at the location
\be
s_{bubble}\sim  \Big ({E\over T}\Big )^{1\over (D-2)}l_p
\label{bubbleapp}
\ee
From (\ref{nnbhp}) we see that hard interactions with Hawking quanta occur at
\be
s_{\alpha}\sim  \Big({E\over T}\Big)^{1\over 2(D-2)}l_p\ll s_{bubble}
\label{nnbhpapp}
\ee
The quantum that falls in from $r-r_0\gtrsim r_0$ has a velocity close to that of light in the local orthonormal frame oriented along the $t, r$ directions. Let us set $t=0$ when the quantum is at the location $s_{bubble}$. We take the metric (\ref{apptwo}), and consider the proper velocity $(U^t, U^r)$  obtained from (\ref{appthree}), (\ref{appseven}): 
\be
U^t=(1-{r_0^{D-3}\over r^{D-3}})^{-1} (1-{r_0^{D-3}\over \bar r^{D-3}})^\h, ~~~U^r\approx -(1-{r_0^{D-3}\over \bar r^{D-3}})^\h
\ee
Thus
\be
{dt\over dr}={U^t\over U^r}\approx -{r_0\over (D-3)}{1\over r-r_0}
\ee
The time to fall between a point $r_i$ to a point $r_f$ is 
\be
\Delta t \approx {r_0\over (D-3)}[ \log( r_i-r_0)-\log( r_f-r_0)]\approx {2r_0\over (D-3)}\log{s_i\over s_f}
\ee
where in the second step we have used (\ref{apps}). In particular the time from the point $s_{bubble}$ to $s_\alpha$ is
\be
\Delta t \sim {r_0\over (D-2)(D-3)}\log \Big ( {E\over T}\Big )
\ee
In all of the above, we have assumed that $E\ll M$, where $M$ is the mass of the hole. But let us for the moment consider the formation of a hole from the collision of two heavy quanta of mass $\sim M$ each, so that $E\sim M$. Then we have (taking the $D=4$ Schwarzschild hole for illustration)
\be
\Delta T \sim M \ln M
\ee
We see that this is of order the scrambling time, when the black hole is expected to stabilize to a thermal form.\footnote{In this analysis we follow the path of the infalling object to see how much Schwarzschild time we have available for tunneling; thus we are following the path of an infalling object.  This computation should be contrasted with the picture depicted in Section 5 of \cite{beyond}, where the spacelike slice is held fixed at small $r$ but evolved to later times at large $r$.}

\refstepcounter{section}
\section*{\thesection \quad Deflection caused by interaction with a $b$ quantum} \label{deflection}

Consider the infalling quantum discussed in Appendix \ref{interact}. We have seen that the interaction probability $P$ becomes order unity at (see Eq.~\ref{aaone})
\be
s\sim  ({E\over T})^{1\over 4} l_p
\label{aaonep}
\ee
At this interaction point, in a local orthonormal frame the infalling quantum has energy (Eq.\;(\ref{aatwo})) 
\be
\omega\sim E{r_0\over s}\sim {E\over T} {1\over s}
\ee
and the Hawking quantum $b$ has energy  (Eq.\;\ref{appthir}))
\be
\omega'\sim {1\over s}
\ee
Let $x$ denote the radial direction (pointing inwards) and let $y$ denote a space direction transverse to the radial direction. We can go to the center-of-mass (cm) frame by a boost along the negative $x$ direction by an amount
\be
\gamma\sim \sqrt{Er_0}\sim \sqrt{E\over T}\gg 1
\ee
where we have assumed that the infalling quantum has $E\gg T$. Since $\gamma\gg 1$, we will set all particle masses to zero. After the interaction, the infalling quantum will have momenta $p_x^{cm}, p_y^{cm}$ in the cm frame. Boosting back to the original frame gives
\be
p_y=p^{cm}_y, ~~~p_x=\gamma p^{cm}_x+v\gamma p^{cm}_t\approx \gamma p^{cm}_x+\gamma\sqrt{(p^{cm}_x)^2+(p^{cm}_y)^2}
\ee
The angle $\theta$ that the infalling quantum makes with the $x$ direction after scattering is give by
\be
\tan\theta={p_y\over p_x}={p^{cm}_y \over \gamma \left( p^{cm}_x+\sqrt{(p^{cm}_x)^2+(p^{cm}_y)^2} \right) }\le {1\over \gamma}\sim ({E\over T})^{-\h}\,.
\label{aathree}
\ee
We are interested in the value of the deflection $\Delta y$ that the quantum will have between the point of interaction and reaching the fuzzball surface. We have
\be
\Delta y =\Delta x \tan\theta 
\ee
The value of $\Delta x$ can be found as follows. The near horizon geometry is Rindler, and can be embedded into Minkowski space as the right Rindler quadrant. The infalling quantum follows a trajectory that is close to an angle $45^0$ in the Minkowski frame.  We want to find $\Delta x$ for the interval between the impact point and the point where the trajectory reaches the fuzzball surface. We find 
\be
\Delta x
= {s\over 2}
\ee
Using  (\ref{aathree}) we get
\be
\Delta y \le {s\over 2}({E\over T})^{-\h}
  \ee
Using (\ref{aaonep}) we get
\be
\Delta y \lesssim l_p ({E\over T})^{1\over 4} ({E\over T})^{-\h}=l_p({E\over T})^{-{1\over 4}}\ll l_p
  \ee
Let us compare this to the wavelength $\lambda_y$ that the infalling quantum had in the $y$ direction. We have assumed that in the rest frame of the infalling detector all quanta have energy $E_{rest}< m_p$. 
Thus
\be
\lambda_y>{1\over m_p}=l_p
\ee
Then for sufficiently large $E/T$, we find
\be
{\Delta y\over \lambda_y}\ll 1 \,,
\ee
that is, the deflection is much smaller than the wavelength of the quanta.

\refstepcounter{section}
\section*{\thesection \quad Fine tuned measurements: dropping in a detector gently} \label{dropping}

In Section \ref{timeqq} we considered the measurements that could be performed on Hawking quanta $b$ when the detector fell in from a point that was {\it not} fine tuned to be close to the horizon. Here we comment briefly on how the discussion changes when we do fine tune to drop the detector in gently from a point close to the horizon. 

Suppose we are interested in detecting quanta of wavelength $\lambda\gtrsim l_p$. These quanta are found at a distance $\bar s \sim \lambda$ from the horizon. To detect these quanta, we need a proper time for switching 
\be
\tau_{available}\gtrsim \lambda\sim \bar s
\label{ten}
\ee
 on the infalling trajectory between the position $\bar s$ and the horizon $s=0$.  
 In \cite{mt} the proper time between these positions was computed for different trajectories parametrized by $\alpha\ge 0$, and it was found that
 \be
\tau_{available}=\bar s e^{-\alpha}\le \bar s
\label{el}
\ee
The maximum value of $\tau_{available}$ is found for $\alpha=0$, which corresponds to dropping in with zero radial velocity from the position $\bar r$. From (\ref{ten}), (\ref{el}), we find that we must drop the detector from rest from a position $\bar s\sim \lambda$. We now face several issues:

\b

(i) Holding the detector at location $\bar s$ before dropping causes it to feel acceleration radiation of wavelength $\bar s\sim \lambda$. So instead of detecting the $b$ quantum that we were interested in, we may pick up spurious quanta from this acceleration radiation. 

\b

(ii) We have to check that the detector quanta are able to interact with the $b$ quanta. We follow the steps in Appendix \ref{interact}, with a few changes. The radiation quantum still has energy given by (\ref{appthir}), so $\omega'\sim{1\over \bar s}$. We let the quanta making the detector have energy $\bar\omega$ each. The probability of interaction is
\be
P\sim {1\over \lambda^{D-2}}G^2 (\omega \bar\omega)^{D-2\over 2}\sim ({\bar\omega\over m_p})^{D-2\over 2}({l_p\over \bar s})^{3(D-2)\over 2}
\ee
where we have used the fact that $\lambda\sim \bar s$. Since $\bar s \gtrsim l_p$, we can get  $P\sim 1$ only if we take $\bar\omega\gtrsim m_p$, which violates our assumption (\ref{four}). 

Detectors are always made, of course, with some amount of internal fine-tuning, so that the detection probabilities are enhanced compared to the simple estimate above. But note that we have constraints in the present situation from the fact that the detector must have a size no more than $\sim \lambda$ and a time of detection that is also no more than $\sim \lambda$. 

\b

(iii) The important issue for us, however,  is the extreme fine tuning involved in lowering the detector with zero velocity to the position $\bar s$: instead of letting the detector fall in from $\bar s\sim r_0$, we have first lowered it to $\bar s\sim \lambda$. This ratio
\be
{\cal R}= {\lambda\over r_0}\ll 1
\label{fift}
\ee
quantifies the required fine-tuning. In our example above, we were interested in detecting TeV quanta, which had a wavelength $\lambda\sim 10^{16} l_p$. To detect $b$ quanta with this energy, we have to first lower our detector to $\bar s\sim 10^{16} l_p$, and then drop it in. The ratio ${\cal R}$ is $\sim 10^{-22}$ for a solar mass hole. Since $\lambda$ is fixed and $r_0$ grows with the size of the hole, ${\cal R}$ is `parametrically small'.  This issue will be important when we come to fuzzball complementarity, since this complementarity is expected to capture non-fine-tuned measurements, but {\it not}  fine-tuned measurements. With ${\cal R}$ satisfying (\ref{fift}), we are not in a situation where a quantum which is `freely falling from afar' impacts the fuzzball surface.

\refstepcounter{section}
\section*{\thesection \quad Passing through the Rindler horizon} \label{rindlercalcs}

In the right Rindler wedge, the Minkowski coordinates $T$, $X$ are related to the Rindler coordinates $t_R$, $x_R$ via
\be
T=r_R \sinh t_R, ~~~X=r_R \cosh t_R \,.
\ee
Let the fuzzball surface be at a distance $\epsilon$ from the horizon. (We will assume $\epsilon\sim l_p$.) Then the Rindler trajectory of a point of the fuzzball surface is
\be
T=\epsilon\sinh{\tau_1\over \epsilon}, ~~~X=\epsilon \cosh {\tau_1\over \epsilon}
\ee
where $\tau_1$ is the proper time along this trajectory. The proper velocity of this point is
\be
U^\mu=(\cosh{\tau_1\over \epsilon}, \sinh {\tau_1\over \epsilon})
\ee
Now consider a particle of mass $m>0$ moving towards this fuzzball surface from the right, with trajectory
\be
T=(\cosh\alpha)   \tau_2, ~~~X=-(\sinh\alpha)   \tau_2+d
\ee
where $\alpha, d$ are constants setting the velocity and initial position of the particle and $\tau_2$ is the proper time along the particle trajectory. The momentum of this particle is
\be
P^\mu=m(\cosh \alpha, -\sinh\alpha)
\ee
 \begin{figure}[htbt]
\begin{center}
\includegraphics[scale=.65]{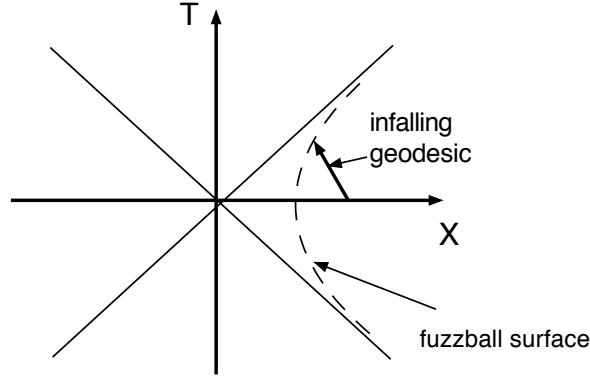}
\end{center}
%
%
\caption{The infalling particle impacting the fuzzball surface.}
\label{fz3p}       
\end{figure}

This particle will impact the fuzzball surface at some point. We wish to compute the energy of this particle in the rest frame of the fuzzball surface, at the moment of this impact (see Fig.\ref{fz3p}). This energy is given by
\be
E'=-P^\mu U_\mu=m(\cosh\alpha \cosh {\tau_1\over \epsilon}+\sinh\alpha \sinh{\tau_1\over \epsilon})\,.
\label{appenergy}
\ee
To find $\tau_1$ at the point of impact, we equate the values of $T, X$ from the two trajectories, getting
\be
(\cosh\alpha) \tau_2 =\epsilon \sinh{\tau_1\over \epsilon}
\ee
\be
-(\sinh\alpha) \tau_2 +d = \epsilon \cosh{\tau_1\over \epsilon}
\ee
The first of these relations gives
\be
\tau_2={\epsilon \sinh{\tau_1\over \epsilon}\over \cosh\alpha}
\ee
Substituting in the second relation gives
\be
\epsilon (\cosh\alpha \cosh {\tau_1\over \epsilon}+\sinh\alpha \sinh{\tau_1\over \epsilon})=d\cosh\alpha
\ee
Using this in (\ref{appenergy}) we get
\be
E'={md\cosh\alpha\over \epsilon}
\ee
Thus we see that we can make $E'$ arbitrarily large by a suitable choice of $\alpha , d$. Choosing large $d$ means we perform the Rindler decomposition with the origin of the Minkowski coordinates far from the present location of the particle; choosing large $\alpha$ means we perform the decomposition in a frame where we have a high relative boost between the two frames. Note that for any $d$, by taking $\alpha$ large we can make $E'$ arbitrarily large, so that the energy of the particle (in the rest frame of the fuzzball surface) is high when it impacts the fuzzball surface.

\end{appendix}

\end{document}